\documentclass[submission,copyright,creativecommons]{eptcs}
\providecommand{\event}{FIT 2010} % Name of the event you are submitting to
\usepackage{breakurl}             % Not needed if you use pdflatex only.

\usepackage{amsthm}

\usepackage{amssymb}
\usepackage[english]{babel}
\usepackage{url}
\usepackage{amsfonts}
\usepackage{stmaryrd}
\usepackage{epsfig}

\usepackage{mathtools}
\usepackage{centernot}

\newcommand{\nota}[1]{}
\newcommand{\opaque}{\square}

\newcommand{\bydef}{\stackrel{\scriptstyle{\it def}}{=}}

\newcommand{\mathaxiom}[2]{
     \begin{array}{l@{\quad}l}%
      {\mbox{\small ({\sc #1})} }
      &{#2}%
     \end{array}}

\newcommand{\mathaxiomtwo}[2]{
   \begin{array}{l}%
    {\mbox{\small ({\sc #1})} }\\[-1mm]
    \quad {#2}%
   \end{array}}

\def \mathrule #1#2#3{\begin{array}{l@{\quad}l}%
    {\mbox{\small ({\sc #1})} }%
    & \irule{#2}{#3}%
\end{array}}

\def \mathruletwo #1#2#3{\begin{array}{l}%
    {\mbox{\small ({\sc #1})} }\\[-.5mm]
    \quad \irule{#2}{#3}%
\end{array}}

\newcommand{\irule}[2]{\frac{\textstyle\rule[-1.3ex]{0cm}{3ex}#1}%
{\textstyle\rule[-.5ex]{0cm}{3ex}#2}}

\def \anmathrule #1#2{
    \irule{#1}{#2}%
}

\newcommand{\outp}[2]{\overline{#1}\langle{#2}\rangle}
\newcommand{\inp}[2]{#1(#2)}
\newcommand{\finp}[2]{{#1}\langle{#2}\rangle}
\newcommand{\ifte}[3]{{\tt if}\ #1\ {\tt then }\ #2\ {\tt else }\ #3}
\newcommand{\fn}{\mathit{fn}}
\newcommand{\bn}{\mathit{bn}}
\newcommand{\vn}{\mathit{vn}}

\newcommand{\ppar}{|\!|}

\newcommand{\tuple}[1]{\widetilde{#1}}

\newcommand{\orbar}{\,\mathrel{\big|}\,}
\newcommand{\as}{\,\mathrel{::=}\,}

\newcommand{\act}{\lambda}
\newcommand{\tr}[1]{\stackrel{#1}{\rightarrowfill}}
\newcommand{\tra}[1]{\stackrel{#1}{\rightarrowtail}}
\newcommand{\ltr}[1]{\stackrel{#1}{\longrightarrow}}

\newcommand{\true}{\mathit{true}}
\newcommand{\false}{\mathit{false}}
\newcommand{\Ev}{\mathit{Ev}}

\makeatletter
\def \rightarrowfill{\m@th\mathord{\smash-}\mkern-6mu%
  \cleaders\hbox{$\mkern-2mu\mathord{\smash-}\mkern-2mu$}\hfill
  \mkern-6mu\mathord\rightarrow}
\makeatother

\makeatletter
\def \Rightarrowfill{\m@th\mathord{\smash-}\mkern-6mu%
  \cleaders\hbox{$\mkern-2mu\mathord{\smash-}\mkern-2mu$}\hfill
  \mkern-6mu\mathord\Rightarrow}
\makeatother

\newcommand{\subj}{\mathit{subj}}
\newcommand{\obj}{\mathit{obj}}
\newcommand{\dom}{\mathit{dom}}
\newcommand{\cod}{\mathit{cod}}

\newcommand{\co}[1]{\overline{#1}}

\newcommand{\nt}[1]{\centernot{\xmapsto{\ #1\ }}}

\title{Contracts for Abstract Processes in Service Composition\thanks{
 Research supported
by the EU FET-GC2 IST-2004-16004 Integrated Project {\sc Sensoria}}}

\author{Maria Grazia Buscemi        
\institute{IMT Lucca Institute for Advanced Studies, Italy} 
        \email{m.buscemi@imtlucca.it}
 \and 
 Hern\'an Melgratti
\institute{FCEyN, University of Buenos Aires, Argentina}
\institute{CONICET}
\email{hmelgra@dc.uba.ar}
}

\def\titlerunning{Contracts for Abstract Processes}
\def\authorrunning{M.G. Buscemi \& H. Melgratti}
\newtheorem{definition1}{Definition}
\newtheorem{proposition}{Proposition}
\newtheorem{theorem}{Theorem}
\newtheorem{lemma}{Lemma}
\newtheorem{example}{Example}

\begin{document}

\maketitle

\begin{abstract}
Contracts are a well-established approach for describing and analyzing behavioral aspects of web service compositions. The theory of contracts comes equipped with a notion of compatibility between clients and servers that ensures that every possible interaction between compatible clients and servers will complete successfully. It is generally agreed that real applications often require the ability of exposing  just  partial descriptions of their behaviors, which are usually known as abstract processes.  We propose a formal characterization of abstraction as an extension of the usual symbolic bisimulation and we recover the notion of abstraction in the context of contracts. 
\end{abstract}

\section{Introduction}
Service Oriented Computing is a paradigm that builds upon the notion of 
services as interoperable elements that can be dynamically discovered through a 
public description
of their interface, which includes their behavior or \emph{contract}.
Session types~\cite{Honda93,DezaniECCOP06,GayHole} and contracts~\cite{LaneveP07,CastagnaGP08,CGP09:TCWS,BZ:TUTCCCC} 
provide a framework for checking whether a client is compliant with a service
and whether a process can be ``safely'' replaced with another one. 
%contracts (e.g., \cite{CastagnaGP08,CGP09:TCWS,BZ:TUTCCCC}) 
Both contracts and session types statically ensure the successful completion 
of every possible interaction between compatible clients and services. 

%Several theories of contracts (e.g., \cite{CastagnaGP08,CGP09:TCWS,BZ:TUTCCCC}) 
%have been
%devised in order to provide a framework for checking whether a client is 
%compliant with
%a service and if a process can be ``safely'' replaced with another one. 
%The use of contracts
%statically ensures the successful completion of every possible 
%interaction between compatible
%clients and services. A key feature in the theory of contracts are 
%\emph{filters}, which are
%coercion functions that are meant to prevent certain behaviors, thus 
%allowing services to
%be compatible with different clients.

In a previous work~\cite{BM:APOL} we have addressed an issue related to 
contracts
by developing a formal theory of \emph{abstract processes} in 
orchestration languages.
An {\em orchestrator} describes the execution flow of a single party in 
a composite
service. The execution of an orchestrator takes control of service 
invocation, handles
service answers and data flow among the different parties in the 
composition. Since
orchestrators are descriptions at implementation level and may contain 
sensitive information
that should be kept private to each party, orchestration comes equipped 
with the notion of
abstract process, which enables the interaction of parties while hiding 
private information.
Essentially, abstract processes are partial descriptions intended to 
expose the protocols
followed by the actual, concrete processes. Typically, abstract 
processes are used for slicing
the interactions of a concrete process over a fixed set of ports. 
%For instance, Consider the following scenario 
%in which an organization sells goods that 
%are produced by a different company. 
As a sample scenario, consider an organization that sells goods that are produced
by another company. The process that handles order requests can be written as follows.
\[
{C_1}\bydef\inp{\it order}{\it desc}.\outp{\it askProd}{\it desc}.\inp{\it
  answProd}{\it cost}.\outp{\it reply}{cost\times1.1}
\]
The process ${C_1}$ starts by accepting an order as a message on port ${\it order}$.  
Then, the received order is forwarded to the actual producer  to obtain a 
quotation. Finally, the client request is answered by sending the production cost 
incremented by a $10\%$. An abstract process of ${C_1}$ should at the same time hide 
the sensitive details of the organization and give enough information to the client for 
allowing interaction. For instance, the following abstract process (where 
$\tau$ stands for a silent, hidden action) shows the interaction of ${C_1}$ with a client.
\[ {A_{C_1}}\bydef\inp{\it order}{\it desc}.\tau.\tau.\outp{\it
  reply}{cost}
\]
%For instance, consider a service 
%$P$ which uses two  ports $a$ and $b$ to communicate 
%with the client while it uses  $c$ to communicate with third-party services.
%An abstraction of $P$ for the client will show only the interactions over $a$ and 
%$b$ but it will hide any action over $c$.  
Another feature of abstract processes is to hide particular values and 
internal decisions made by concrete processes. Consider, e.g., the following 
process for authorizing loans.
\begin{small}
\[
{C_2}\bydef\inp{\it request}{\it amount,salary}.\ifte{( salary  > amount / 50)}{\outp{\it refuse}{}}{\outp{\it approved}{}}
\]
\end{small}
 Suppose also that the bank does not want to publicly declare its policy, under which 
 a loan is approved only when the requested amount is at most 50 times the solicitor's salary. 
 This can be achieved by providing an abstract process where some values are
 {\it opaque} (noted with $\opaque$), i.e., not specified. An abstract process of ${C_2}$ can be as below.
\[
 {A_{C_2}}\bydef\inp{\it request}{\it amount,salary}.\ifte{salary > \opaque}
  {\outp{\it refuse}{}}{\outp{\it approved}{}}
\]
Note that the conditional process in ${A_{C_2}}$ has to be thought of as an
internal, non-deterministic choice in which the bank may decide either
to approve or to refuse the application. In other words, the client cannot
infer from  ${A_{C_2}}$  the actual decision that the bank will take. In general, we 
require an abstraction to provide enough information to decide whether a client and a 
service are compliant, i.e., whether their interaction will allow them to complete their 
execution or not.

In~\cite{BM:APOL} we have characterized the valid abstractions of a 
concrete orchestration and we have shown that valid abstractions preserve
compliance. More precisely, we have formally defined suitable abstractions of concrete processes
as a relation among abstract and concrete processes, called \emph{simulation-based abstraction} relation,
which is an extension of the usual symbolic bisimulation \cite{HL:SB,BD:SSPC}. 

A main goal of the present paper is to investigate the relation between simulation-based abstraction 
and contracts. In particular, we aim at recovering the notion of abstraction in the context of 
the theory of contracts developed in~\cite{CGP09:TCWS}. 
Contracts are types describing the external, visible behavior of a service. Contracts come 
equipped with a notion of service compatibility that characterizes all the valid clients of a service, i.e., 
the clients that terminate any possible interaction with the service. 
In this sense, contracts can be used to statically ensure that the composition of two services
is safe. Contract compatibility induces a preorder relation ($\prec$) among contracts that characterizes 
the safely replacement of services. For instance,  considering two contracts $\sigma_1$ and $\sigma_2$, if 
$\sigma_1\prec\sigma_2$ we know that any valid client of $\sigma_1$ is also a valid client of $\sigma_2$,
hence $\sigma_1$ can be substituted by $\sigma_2$ in any context. 
A contract for the service ${C_1}$ corresponding to the selling company example introduced above can be 
written as follows. 
\[
{\sigma_1}\bydef{\it order}.\overline{\it askProd}.{\it answProd}.\overline{\it reply}
\]

Note that $\sigma_1$ describes the interactions of $C_1$ with both the client and the producer. 
Hence, we would like to use the idea of abstraction in the context of contracts to obtain
slices of the behaviour of a service and to reason about the interactions of a 
service with a particular partner, i.e., we would like to use $\sigma_1$ to conclude that 
any client behaving as $\rho_1 = \overline{\it order}.{\it reply}$ is compliant with the role 
client of the service $\sigma_1$. 

More in detail, given a contract $\sigma$  and a role, defined in terms of a set of visible actions $V$, 
the abstraction $\mathcal{A}_V(\sigma)$ of $\sigma$ can be thought as the contract that hides all the actions
in $\sigma$ that do not
appear in $V$. For instance, the abstraction of $\sigma_1$ for the role client will be as follows
\[
\mathcal{A}_{\{order, reply\}} (\sigma_1) \bydef {\it order}.\overline{\it reply}
\]
Another key property of the abstraction type is to turn guarded choices into internal choices, if some guards are hidden. For instance, consider the process $P \equiv \inp a{}.\overline c \langle \rangle + \inp b{}.\overline d \langle \rangle$. The type of $P$ is $\sigma = a.\overline c + b. \overline d$. If we hide $a$, the abstraction type of $\sigma$ is
\[ 
\mathcal{A}_{\{b,c,d\}} (\sigma) \bydef \overline c \oplus b. \overline d 
\]
The main technical contributions of this work are the following. Firstly, we define abstraction as a function $\mathcal{A}_V(\_)$ over contracts and show that our definition 
behaves well with respect to safe replacement, i.e.,  $\mathcal{A}_V(\sigma)$ can be substituted by $\mathcal{A}_V(\rho)$ 
whenever $\sigma$ can be substituted by $\rho$. Technically speaking, this fact amounts to proving that 
$\sigma \prec \rho$ implies $\mathcal{A}_V(\sigma)\prec \mathcal{A}_V(\rho)$,  when taking 
$\prec$ as the strong subcontract preorder.

Then, we show that contract abstraction can be used on top of contracts to reason about
 slices of a concrete service. That is, given a suitable type system for assigning 
contracts to concrete services, the type of a particular slice of a concrete 
service can be defined simply as the abstraction of the original contract. Formally, we 
show that any consistent type system enriched with a typing rule that assigns 
any slice of a concrete service with 
the corresponding contract abstraction is a consistent type system. This result allows us 
to use abstraction over contracts to reason about slices of a concrete service, e.g., we 
can use $\mathcal{A}_{\{order, reply\}} (\sigma_1)$ to safely reason about the interactions
of $C_1$ with a client. 
 
Finally,\   we show that contract abstraction matches simulation-based abstraction. Consider the 
simu-lation-based abstraction $Q$ 
of $P$ that characterizes a particular slice of $P$. Assume that $Q$ has contract $\sigma$ and 
consider any compliant client $C$ of $Q$ (namely, the type of $C$ is compliant with $\sigma$). Our 
results ensure that 
$C$ is also compliant with the slice of $P$ described by $Q$. 

%First, we provide a type system for concrete and abstract orchestrators that associates
%each orchestrator with a contract as its type. Our type system resembles the one proposed in~\cite{CGP09:TCWS}
%for typing web service orchestrations. The main difference is that in our case we also consider
% orchestrators performing silent actions and 
%hidden decisions. Then, .

\section{Abstract processes}
\label{sec:concrete-processes}

In this section we recall the language of abstract processes proposed
in~\cite{BM:APOL} along with a notion of abstraction relation over processes, 
which is a generalisation of \cite{BD:SSPC}.
%The computation model we present is highly inspired by the
%composition model of {\sc ws-bpel}, which can be roughly described as
%follows:  a composite service can be though as the 
%parallel composition of several orchestrators that interact by 
%exchanging XML-documents using one of a set of basic 
%actions. 
First, we introduce the language of {\em abstract processes}, 
which is a version of value-passing {\sc  ccs}~\cite{Mil:CC} with 
input guarded choices and conditional statements but without recursion 
plus the possibility of having {\em opaque} definitions.
%We start by defining \emph{concrete} processes and 
%then extend to \emph{abstract} processes, which are given by using 
%the primitives of the concrete processes plus the possibility of having 
%{\em opaque} definitions. 
An opaque element is meant to hide the precise value of an element: for 
instance, an opaque assignment to a data variable hides the assigned value. 
We assume the set of data values to be finite so that the present 
version of the calculus can be encoded into the fragment without 
value-passing. We refer the interested reader to \cite{Mil:CC} for a more 
detailed treatment. 

\paragraph{Syntax}
\label{sec:synt-inform-semant-concrete}
We assume an infinite denumerable set of names $\mathcal{N}$
that is partitioned into a set of port names $\mathcal{X}$,
a set of finite data variables $\mathcal{V}$,
and a finite set of data constants $\mathcal{C}$.
We write the special name $\opaque$ to denote an opaque element, and 
we assume $\opaque\not\in\mathcal{N}$. We let $\eta$ range over 
$\mathcal{N}\cup\opaque$, $u,v,\ldots$ range over $\mathcal{V}$, 
$a,b,c,\ldots$ range over $\mathcal{C} \, \cup \, \{\opaque\}$, and 
$x,y,z,\ldots$ range over $\mathcal{X}$. We let $m,n,\ldots$ 
range over $\mathcal{V} \,\cup\, \mathcal{C} \, \cup \, \{\opaque\}$. 
We write $\tilde{\eta}$ for a tuple of names. {\em Substitutions}, 
ranged over by $\sigma$, are partial maps from 
$\mathcal{V}$ onto $\mathcal{V} \,\cup\, \mathcal{C} \, \cup \, \{\opaque\}$. 
Domain and co-domain of $\sigma$, noted $\dom(\sigma)$ and $\cod(\sigma)$, are 
defined as usual.  By $m \sigma$ we denote $\sigma(m)$ if $m \, \in 
\dom(\sigma)$, and $m$ otherwise. 

The set of {\em abstract processes} $P$ is given by the following grammar:
\[
  P \;\; \as \;\; 0 \orbar P \,| \, P
     \orbar \tau.P
     \orbar \outp{x}{\tuple m}.P
     \orbar \inp{x_1}{\tuple{v_1}}.P+\ldots+\inp{x_n}{\tuple{v_n}}.P
     \orbar \ifte{m=n}P P
\]
As usual, $0$ stands for the inert process, $P\,|\,P$ for the 
parallel composition of processes, 
$\tau.P$ for the process that performs a silent action and then behaves 
like $P$, $\outp{x}{\tuple{m}}.P$ for the process that sends the message 
$m$ over the port $x$ and then becomes $P$. The process
$\inp{x_1}{\tuple{v_1}}.P_1+\ldots+\inp{x_n}{\tuple{v_n}} .P_n$
denotes an external choice in which some process $\inp{x_i}
{\tuple{v_i}}.P_i$ is chosen when the corresponding guard $\inp{x_i}
{\tuple{v_i}}$ is enabled. The conditional process $\ifte{m=n}P P'$
behaves either as $P$ if $m$ and $n$ are syntactically equivalent, or
as $P'$ otherwise. Opaque names can appear either as subjects of 
input and output prefixes, values of output prefixes, or parts of 
conditions in $\ifte{\_}{\_}{\_}$ processes, but not as a bound variables. 
A conditional statement becomes an \emph{internal} choice when 
at least one value in the condition is opaque; similarly, a guarded choice  
becomes an internal choice when the subject of the input guard is the opaque 
name. 

We let $P,Q,R \ldots$ range over abstract processes and we simply write 
process to denote an abstract process. By \emph{concrete} processes we 
denote processes not containing opaque names. Note that in 
$\inp{x_1}{\tuple{v_1}}.P_1+\ldots+\inp{x_n}{\tuple{v_n}}.P_n$, the data 
variables $v_i$ are bound, for all $i$. We use the standard notions 
of {\it free} and {\it bound} names of processes, noted respectively as 
$\fn(P)$ and $\bn(P)$, and $\alpha$-conversion on bound names. We assume 
that the sets of free and bound names are disjoint 
and that the bound names of a process are all distinct from each other. 
As usual, a process $P$ is {\em closed} if $\fn(P) \cap \mathcal{V} =
\emptyset$. We also adopt the usual convention of omitting trailing $0$'s. 

\subsection{Symbolic semantics}
%The operational semantics, as usual, is given in two steps: the
%definition of a {\em structural congruence}, which rearranges 
%processes into adjacent positions, and a notion of {\em 
%labelled transition relation} that captures computation on processes.  
For the purpose of this paper we only recall the \emph{symbolic} labeled 
transition relation over processes, 
%which is in the spirit of~\cite{HL:SB,BD:SSPC}
while we report in Appendix \ref{app:non-symb} the non-symbolic semantics along with a proof that the two semantics are \
equivalent.
%In the labeled transition rules, labels have two components: a 
%symbolic action $\act$ and a Boolean condition $M$ over
%the set of data variables and data constants $\mathcal{V} \,\cup\,
%\mathcal{C}$ that must hold for the $\alpha$-transition to be enabled.

We define {\em structural congruence}, $\equiv$, as the least
congruence over processes that is closed with respect to $\alpha$-conversion 
and such that the set of process is a monoid with respect to parallel 
composition $|$ (being $0$ the neutral element). 
%the following rules:
%\[
%\begin{array}{r@{\;\equiv\;}l@{\hspace{5mm}}r@{\;\equiv\;}l@{\hspace{5mm}}r@{\;\equiv\;}l}
%  P | 0 & P
%& P_1 | P_2 & P_2 | P_1
%& (P_1 |P_2)|P_3 & P_1 |(P_2|P_3)
%\end{array}
%\]

We let {\em symbolic actions} $\act$ range over the {\em silent move}, 
{\em input} and {\em free output} and we let {\em conditions} $M$ range 
over a language of Boolean formulas:
\[
  \act ::= \tau \;|\; \inp x {\tuple v} \;|\;\outp x {\tuple m}
 \qquad 
  M \; ::= \; \true \;|\; \false \;|\; m = n \;|\; m \neq n \;|\; 
  M \wedge M \;|\; M \vee M.  
\]
As usual, for $\act \neq \tau$, $\subj(\act)$ and $\obj(\act)$ 
denote the subject and the object of $\act$ respectively.  
The notions of {\it free} names $\fn(\cdot)$, {\it bound} names $\bn(\cdot)$,
and $\alpha$-conversion over actions and conditions are as expected,
considering that the occurrences of the names $v_i$'s are bound in
$\inp x {\tuple v}$ and that conditions have no bound names.
%% controllare se queste tre righe sotto servono
For $X$ a process or an action, $X\sigma$ denotes the expression obtained
by replacing in $X$ each data variable $u \in \fn(X)$ with $u \sigma$, 
possibly $\alpha$-converting to avoid name capturing.
By $M \sigma$ we mean the condition obtained by simultaneously
replacing in $M$ each data variable $v \in fn(M)$ with $v \sigma$.  A
condition $M$ is {\em ground} if $M$ does not contain data variables.
The {\em evaluation} $\Ev(M)$ of a ground condition $M$ into the set
$\{ \true, \false \}$ is defined by extending in the expected
homomorphical way the following clauses:
\[
 \begin{array}{l@{\;\;\;}l@{\;\;\;}l}
  \Ev(\true) = \true & \Ev(a=a) = \true & \Ev(a=b)= \true \;\; \mbox{if $\{ a,b\} \cap {\opaque} 
  \neq \emptyset$} \\
  \Ev(\false) = \false & \Ev(a=b) = \false \;\; \mbox{if $a,b \neq \opaque$} 
  & \Ev(a \neq b)= \true \;\; \mbox{if $\{ a,b\} \cap {\opaque} 
  \neq \emptyset$}
 \end{array}
\]
A substitution $\sigma$ {\em respects} $M$, written $\sigma \models
M$, if $M\sigma$ is ground and $\Ev(M\sigma)=\true$. A condition $M$
is \emph{consistent} if there is a substitution $\sigma$ such that
$\sigma \models M$.  A condition $M$ {\em logically entails} a
condition $N$, written $M \Rightarrow N$, if, for every $\sigma$,
$\sigma \models M$ implies $\sigma \models N$.  
%$M$ and $N$ are {\em logically equivalent}, written 
%$M \Leftrightarrow N$, if $M \Rightarrow N$ and $N \Rightarrow M$.  
For instance, $v = a \wedge u \neq b \wedge v = u \Rightarrow a \neq b$ 
and $\true \Rightarrow u = a \vee u \neq a$.  For $\lambda$ a symbolic 
action and $\sigma$ a substitution such that every data variable in 
$\lambda$ belongs to $\dom(\sigma)$, we write $\lambda \sigma$ to 
denote the following action:
 \[
   \act \sigma \mathrel{\mathop{=}\limits^{\rm def}}   
     \left \{ 
             {\begin{array}{l@{\qquad}l}
              \tau & \mbox{if $\lambda = \tau$}\\[1mm]
              \outp x {a_1,\ldots,a_k} & \mbox{if $\lambda = \outp x 
               {n_1,\ldots,n_k}$ and $a_i = n_i \sigma$ for 
                $i=1,\ldots,k$}\\[1mm]
              \finp x {a_1,\ldots,a_k} & \mbox{if $\lambda = \inp x 
                {v_1,\ldots,v_k}$ and $a_i = \sigma(v_i)$ for $i=1,\ldots,k$}
              \end{array}}
      \right .
 \]
By $\act = \act'$ we denote the following condition:
\[
 \act = \act' \mathrel{\mathop{=}\limits^{\rm def}}
   \left \{ 
             {\begin{array}{l@{\qquad}l}
   \true & \mbox{if $\act=\act'=\tau$ or $\act= \act' = \inp x {\tuple v}$}\\  
   \tuple m  = \tuple n & \mbox{if $\act= \outp x {\tuple m}$ and 
    $\act'=\outp x {\tuple n}$}\\
  \false & \mbox{otherwise}
  \end{array}}
 \right .
\]  
For $M$ a condition and $D =\{M_1,\ldots,M_n\}$ a finite set of
conditions, $D$ is a {\em $M$-decomposition} if $M\Rightarrow M_1 \vee
\ldots \vee M_n $.  For instance, $\{u=a, u \neq a \}$ is a
$\true$-decomposition.

The symbolic labeled transition relation $\tr {M, \act}$ over abstract 
processes is the least relation satisfying the inference rules in
Table~\ref{table:lts}. 
%The additional symbolic rules for processes are given in Table \ref{table:lts-abstract}. 
Intuitively, the condition $M$ in the label $M, \act$ of a transition 
collects the Boolean constraints on the free data variables of the source 
process necessary for action $\act$ to take place. For instance, the rules 
for prefixes say that each prefix can be consumed unconditionally, while 
rules ({\sc s-if}) and ({\sc  s-else}) make the equalities or inequalities of 
the conditional statements explicit. For instance, the process $P \equiv \inp x v.
\ifte{v =a} {\outp y v}{0}$, after a first step, can make a transition
under condition that variable $v$ is equal to $a$:
\[
  P \tr {\true, \inp x v} \ifte{v =a} {\outp y v}{0} \tr {v = a, \outp y v} 0 
\]
As another example, consider the process $R \equiv \inp{\opaque}\tau.P+\inp{x}{v_2}.Q$. 
By rule ({\sc s-choice-3}), a possible move for $R$ is $R \tr{\true, \inp x {v_2}} Q$, 
where the input guard is executed. Another possibility is $R \tr{\true,\tau} P$, where 
$R$ makes an internal choice. 
%Note that in rule ({\sc s-in}) input variables are 
%{\em not} instantiated immediately. 

%Remark that the present rules are simpler than those given in
%\cite{BD:SSPC} for the pi-calculus because our calculus is a
%value-passing {\sc ccs} plus conditional statements and, thus, logical
%conditions do not affect channel names.

%\begin{proposition}
% Let $P$ be a closed process. 
% \begin{itemize}
% \item If $P \tr {\alpha} Q$ then there exist $R$, $M$, $\act$, and
%   $\sigma \models M$ s.t. $P \tr {M, \act} R$, $\alpha = \act
%   \sigma$ and $Q = R \sigma$.
% \item If $P \tr {M, \act} Q$ then there exists $\sigma \models M$
%   such that $P \tr \alpha Q \sigma$ and $\alpha = \act \sigma$.
% \end{itemize}
%\end{proposition}
%
\begin{table}[t]
 \[
\begin{array}{l}
  \mathaxiomtwo{s-tau}
		{\tau.P \tr {\true, \tau} P}
\hfill
  \mathaxiomtwo{s-out}
		{\outp{x}{\tuple{m}}.P \tr {\true, \outp{x}{\tuple{m}}} P}
\hfill
  \mathaxiomtwo{s-in}
  {\inp{x_1}{\tuple{v_1}}.P_1+ \ldots+\inp{x_n}{\tuple{v_n}}.P_n
  \tr {\true, \inp{x_i}{\tuple{v_i}}} P_i}
\\[5pt]
\mathrule{s-par}
					 {P\tr {M, \act} P' \quad \bn(\act) \cap \fn(Q) = \emptyset}{P\;|\;Q\tr{M,\act}P'\;|\;Q}
                                         \hfill
                                         \mathrule{s-if}
                                         {P\tr {M, \act} P' \quad m=n \wedge M~\mbox{consistent}}
                                         {\ifte{m=n}{P}{Q} \tr {m=n \wedge M,\act} P'}
                                         \\[15pt]
                                         \mathrule{s-str}
                                         {P\equiv Q\quad Q\tr{M, \act} Q'\quad Q'\equiv P'}
                                         {P\tr{M, \act} P'}
                                         \quad
                                         \mathrule{s-else}
					 {Q\tr {M, \act} Q' \quad m\neq n \wedge M~\mbox{consistent}}
                                         {\ifte{m=n}{P}{Q} \tr {m\neq n \wedge M,\act} Q'}
                                         \\[15pt]
                                           \mathruletwo{s-choice-1}
  {P \tr {M, \act} P' \quad  \opaque\in\{m,n\}}
  {\ifte{m=n}{P}{Q} \tr {M, \act} P'}
  \qquad \qquad 
  \mathruletwo{s-choice-2}
  {Q \tr {M, \act} Q' \quad  \opaque\in\{m,n\}}
  {\ifte{m=n}{P}{Q} \tr {M, \act} Q'}
  \\[25pt]
  \mathaxiom{s-choice-3}
  {\inp{x_1}{\tuple{v_1}}.P_1+
  \ldots+\inp{\opaque}{\tuple{v_i}}.P_i+\ldots+\inp{x_n}{\tuple{v_n}}.P_n
  \tr {\true, \tau} P_i}
\end{array}
\]
\caption{Symbolic LTS for processes}
\label{table:lts}
\end{table}

\paragraph{Non-symbolic semantics.} \
%We define a non-symbolic semantics of the calculus as below. 
The following definition corresponds to the original semantics proposed in~\cite{BM:APOL}. (Details are in Appendix \ref{app:non-symb}.)

\begin{definition1}[Non-symbolic semantics] \label{def:alt-non-symb-sem}
Let $P$, $Q$ and $\lambda$ be closed terms. $P\xrightarrow{\lambda} Q$ iff  $P\xrightarrow{M,\lambda'} P'$, $\sigma\models M $, $\lambda = \lambda'\sigma$ and $Q = P'\sigma$.
\end{definition1}

\subsection{Simulation-based abstraction}

\begin{definition1}[visible names] Given a set of visible names $V$ and
a symbolic action $\act$, the set of visible received names of $\act$, written
$\vn({\act})_V$, is defined as follows:
   \[
   \vn({\act})_V \mathrel{\mathop{=}\limits^{\rm def}}   
     \left \{ 
             {\begin{array}{ll}
              \tuple{u} & \mbox{if $\act = \inp x {\tuple u}$ and $x\in V$}\\[1mm]
              \emptyset & \mbox{otherwise}\\ 
              \end{array}}
      \right .
 \]
We will omit the subscript $V$ when it is clear from the context.
\end{definition1}

\begin{definition1}[simulation-based abstraction]\label{def:abstraction}
 The family $\mathcal{R}= \{\mathcal{R}^{V}_{\;M}\}_M$ of process 
 relations is a {\em family of simulation-based abstraction relations},
 indexed over the set of conditions $M$, iff for all $M$ and 
 $P \mathcal{R}^{V}_{\;M} Q$:  
 \begin{enumerate}
 \item If $Q \tr {N, \act} Q'$ and $\bn(\act)\cap\fn(P,Q,M) =\emptyset$
   then there exists a $M\wedge N$-decomposition $D$ s.t. $\forall 
   M'\in D$ there exists $P \tr {N',\act'} P'$, with
   $M'\Rightarrow N' \wedge \act_{|V} = \act'$ 
   and $P' \mathcal{R}^{V \cup \vn(\act) }_{\;M'} Q'$.
 \item if $P \tr {N, \act} P'$ and $\bn(\act)\cap\fn(P,Q,M) =\emptyset$
   then there exists a $M\wedge N$-decomposition $D$ s.t. $\forall 
   M'\in D$ there exists $Q \tr {N',\act'} Q'$ with $M'\Rightarrow 
   N'_{|V} \wedge \act = \act'_{|V}$ 
   and $P' \mathcal{R}^{V \cup \vn(\act') }_{\;M'} Q'.$  
 \end{enumerate}
 A process $P$ is a {\em simulation-based abstraction} of a process
 $Q$ with respect to a set $V \subseteq \mathcal{N}$, written $P
 \propto^V Q$, if there is an abstraction relation
 $\mathcal{R}^{V}_{\;\,\true}$ s.t. $P \, \mathcal{R}^{V}_{\;\,\true}
 \,Q$, with $\fn(P) \subseteq V$.
\end{definition1}
Condition 1 above states that the abstraction $P$ simulates the
concrete process $Q$ up to hidden names. Note that we require 
$\act_{|V} = \act'$ instead of the standard definition of symbolic 
bisimulation that imposes the exact matching of action labels.  
Condition 2 states that the (concrete) process $Q$ can simulate its abstraction 
$P$ if we forget about the constraints involving hidden values. 
That is, if $P$ proposes a
move with label $\langle N, \act \rangle$ we allow $Q$ to mimic the
behavior for a more restrictive condition $N'$. (Actually, $N'$ 
may contain several additional constraints involving hidden names.) 
Note that this makes the abstraction relation not symmetric.
For instance, consider the two processes below:
 \[
   P \equiv \ifte {v=\opaque} {\outp y v} {\outp z v} \quad 
   Q \equiv \ifte {v=a} {\outp y v} {\outp z v}. 
 \]
 It holds that $P\propto^V Q$ for $V = \{v,y,z\}$. Indeed, when
 considering the transition $P \tr {\true,\outp y v} 0$, we can take
 $Q \tr {v=a,\outp y v} 0$ since $\true \Rightarrow (v=a)_{|V} \wedge
 \outp y v = \outp y v_{|V}$.  Conversely, $P\not\propto^V Q'	
 \equiv \ifte {a=a} {\outp y v} {\outp y v}$ because $P
 \tr {\true,\outp z v} 0$ but $Q' \not\tr {M,\outp z v}$. 
  We remark that the relation $\propto$ is a simulation (since the abstract
 process simulates the concrete one) but, in general, is not either 
 a bisimulation or a similarity.

\section{Theory of contracts}

This section summarizes the basics about the theory of contracts proposed 
in~\cite{CastagnaGP08,CGP09:TCWS}. Let $\mathcal{N}$ be a set of names, the set of contracts $\Sigma$ is 
given by the following grammar.

\begin{math}
   \begin{array}{l@{\ ::=\ }ll}
     \alpha & a \ |\ \co{a} & a\in\mathcal{N}\\
     \sigma & 0 \ |\ \alpha.\sigma \ |\ \sigma\oplus\sigma \ | \ \sigma+\sigma
   \end{array}
\end{math}

The contract 0 describes a service that does not perform any action. The contract
$\alpha.\sigma$ stands for a service that is able to execute $\alpha$ and then 
continues as $\sigma$. The contract $\sigma + \rho$ describes a service that lets
the client decide whether to continue as $\sigma$ or as $\rho$, while $\sigma\oplus\rho$ stands for a service that internally decides whether to continue
as $\sigma$ or $\rho$. As usual, trailing $0$'s are omitted. Contracts will be considered modulo associativity of each sum operator. We 
usually write summations 
$\sigma_1+\sigma_2+\ldots+\sigma_n$ and $\sigma_1\oplus\sigma_2\oplus\ldots\oplus\sigma_n$ respectively as  $\Sigma_{i\in\{1,\ldots,n\}} \sigma_i$ and $\bigoplus {i\in\{1,\ldots,n\}} \sigma_i$. By convention, $\Sigma_{i\in\emptyset} \sigma_i = 0$.

In this paper we restrict our attention to finite contracts, although the presentation in~\cite{CGP09:TCWS} deals also with infinite contracts in the form of infinite trees that satisfy regularity and a contractivity condition.

The operational semantics of contracts is given in terms of the \textsc{lts} defined
below.

\begin{definition1}[Transition] Let $\sigma \nt{\alpha}$ be the least relation such that:

\[
   \begin{array}{l@{\qquad}l@{\qquad}l@{\qquad}l}
      0 \nt{\alpha}
      &
      \anmathrule{\alpha\neq \beta}{\beta.\sigma \nt{\alpha}}
      &
      \anmathrule{\sigma\nt{\alpha}\quad \rho\nt{\alpha}}{\sigma\oplus\rho\nt{\alpha}}
      &
      \anmathrule{\sigma\nt{\alpha}\quad \rho\nt{\alpha}}{\sigma+\rho\nt{\alpha}}\\
   \end{array}
\]

The transition relation of contracts, noted $\xmapsto{\alpha}$, is the least
relation satisfying the rules

\begin{math}
  \begin{array}{l@{\;\;\;}l@{\;\;\;}l@{\;\;\;}l@{\;\;\;}l}
     \alpha.\sigma \xmapsto{\alpha} \sigma
     &
     \anmathrule{\sigma\xmapsto{\alpha} \sigma' \quad \rho\xmapsto{\alpha}\rho'}
                {\sigma+\rho\xmapsto{\alpha} \sigma'\oplus\rho'}
     &
     \anmathrule{\sigma\xmapsto{\alpha} \sigma' \quad \rho\nt{\alpha}}
                {\sigma+\rho\xmapsto{\alpha} \sigma'}
     &
     \anmathrule{\sigma\xmapsto{\alpha} \sigma' \quad \rho\xmapsto{\alpha}\rho'}
                {\sigma\oplus\rho\xmapsto{\alpha} \sigma'\oplus\rho'}
     &
     \anmathrule{\sigma\xmapsto{\alpha} \sigma' \quad \rho\nt{\alpha}}
                {\sigma\oplus\rho\xmapsto{\alpha} \sigma'}
  \end{array}
\end{math}

and closed under mirror cases for the external and internal choices. 
\end{definition1}

The operational semantics for contracts handles choices differently from the standard \textsc{ccs} transition system. Traditional \textsc{ccs} rules for a 
choice commits to the execution of a branch as soon as it performs the first 
action of the branch, e.g., $a.b + a.c$ reduces  to both $b$ and $c$. 
Differently,  the contract $a.b + a.c$ has only the continuation $b\oplus c$, i.e., the operational
semantics does not provide any information about the actual choice that has been 
taken, in this way the environment is aware of the fact that the system will 
internally decide whether to behave as $b$ or $c$. Consequently, for any 
action $\alpha$ and contract $\sigma$ there is at most one contract $\sigma'$ 
such that $\sigma\xmapsto{\alpha}\sigma'$. Let $\sigma\xmapsto{\alpha}\sigma'$, 
we write $\sigma(\alpha)$ for the unique continuation of $\sigma$ after $a$ (i.e.,
$\sigma(\alpha) = \sigma'$). We use $\mathtt{init}(\sigma)$ to denote the 
set of actions that can be immediately emitted by $\sigma$, i.e., $\mathtt{init}(\sigma)=\{\alpha\ |\ \exists \sigma' s.t. \sigma \xmapsto{\alpha} \sigma'\}$.

\begin{definition1}[Ready sets] Let $\mathcal{P}_f(\mathcal{N}\cup\co{\mathcal{N}})$ be
the set of finite parts of $\mathcal{N}\cup\co{\mathcal{N}}$, called \emph{ready sets}. Let also $\sigma\Downarrow R$ be the least relation between contracts 
$\sigma \in \Sigma$ and ready sets $R$ in $\mathcal{P}_f(\mathcal{N}\cup\co{\mathcal{N}})$ such that

\begin{math}
\begin{array} {l}
   0\Downarrow \emptyset
   \quad
   \alpha.\sigma\Downarrow \{\alpha\}
   \quad
   \anmathrule {\sigma\Downarrow R \quad \rho\Downarrow S}
               {\sigma+\rho \Downarrow R \cup S} 
   \quad
   \anmathrule {\sigma\Downarrow R}
               {\sigma\oplus\rho \Downarrow R} 
   \quad
   \anmathrule {\rho\Downarrow R}
               {\sigma\oplus\rho \Downarrow R} 
\end{array}
\end{math}
\end{definition1}

As usual we make $\co{\co{a}}=a$. For a given ready set $R$, $\mathtt{co}(R)$ stands for its complementary ready set, i.e., $\mathtt{co}(R) = \{\co{\alpha}\ |\ \alpha\in R\}$. 

\subsection{Compliance and subcontract relation} 

Compliance formally states when the behavior of a client complies with the 
behavior of a service. It is assumed that the behavior of both the client and 
the service are described by contracts. There is a reserved special action $\mathtt{e}$ (for ``end'') that can occur in client contracts and that represents
the ability of the client to successfully terminate. Compliance requires that,
whenever no further interaction is possible between the client and the service, the 
client be in a state where this action is available. 

\begin{definition1}[Strong compliance]
\label{def:strongcompliance}
 $\mathcal{C}$ is a \emph{strong compliance} relation
if $(\rho,\sigma) \in \mathcal{C}$ implies that 
\begin{enumerate}
	\item $\rho\Downarrow R$ and $\sigma\Downarrow S$ implies either $\mathtt{e}\in R$ or $\mathtt{co}(R)\cap S \neq \emptyset$, and 
	\item $\rho\xmapsto{\co{\alpha}} \rho'$ and $\sigma\xmapsto{\alpha} \sigma'$
	implies $(\rho',\sigma')\in \mathcal{C}$.
\end{enumerate}

We use $\dashv$ to denote the largest strong compliance relation.
\end{definition1}

Once the precise notion of compliance between clients and services has been established, the notion of  strong subcontract is defined. A contract $\sigma$
is a strong subcontract of another contract $\rho$ when all clients compliant with
$\sigma$ are also compliant with $\rho$. This notion is coinductively defined as follows. 

\begin{definition1}[Strong subcontract] $\mathcal{S}$ is a \emph{strong subcontract relation}
if $(\sigma, \rho)\in\mathcal{S}$ implies that
\begin{enumerate}
	\item $\rho\Downarrow R$ implies that there exists $S\subseteq R$ such that $\sigma\Downarrow S$, and
	\item $\rho\xmapsto{\alpha}\rho'$ implies $\sigma\xmapsto{\alpha}\sigma'$ and $(\sigma', \rho')\in\mathcal{S}$.
\end{enumerate}

We denote with $\sqsubseteq$ the largest strong subcontract relation.

\end{definition1}

It has been shown in~\cite{LaneveP07} that $\sqsubseteq$ is the \emph{must testing preorder} as defined by~\cite{dNH:TEP}.

\subsection{Assigning contracts to ordinary processes}
Contracts are intended as types for describing the behavior of concrete implementations. It is assumed that the observable behavior of concrete implementations is described by a labeled transition so that $P\xrightarrow{\mu} P'$
describes the evolution of a process P that performs an  action $\mu$ and then becomes
$P'$. The performed action $\mu$ can be either a visible action (e.g., an input $a$ or an output $\co{a}$) or an internal, invisible action $\tau$ that the process 
$P$ executes autonomously. Then, it is assumed that clients and servers interact by synchronizing over complementary actions, as 
it is formally stated below.

\begin{definition1}[Strong process compliance] Let $P\ppar Q \rightarrow P'\ppar Q'$ be the least relation defined by the rules:
\begin{math}
  \anmathrule{P\xrightarrow{\tau}P'}{P\ppar Q \rightarrow P'\ppar Q}
  \quad
  \anmathrule{Q\xrightarrow{\tau}Q'}{P\ppar Q \rightarrow P\ppar Q'}
  \quad
  \anmathrule{P\xrightarrow{\alpha}P'\quad Q\xrightarrow{\co{\alpha}}Q'}{P\ppar Q \rightarrow P'\ppar Q'}
\end{math}
\end{definition1} 

The reflexive and transitive closure of $\rightarrow$ is written $\Rightarrow$;
$P\ppar Q \rightarrow$ stands for  $P\ppar Q \rightarrow P'\ppar Q'$ for some $P'$
and $Q'$. We write $P\ppar Q \nrightarrow$ if not $P\ppar Q \rightarrow$.
 A computation of $P\ppar Q$ is maximal if either it is infinite or there exists
$P_n\ppar Q_n$ such that $P\ppar Q \Rightarrow P_n\ppar Q_n \nrightarrow$.
The client $P$ is strongly compliant with the service Q, written $P\dashv Q$, 
if for every configuration $P_i\ppar Q_i$ of every maximal computation there 
exists $j \geq i$ such that either $P_j \xrightarrow{\alpha} P_{j+1}$ for some
$\alpha$ or $P_j \centernot{\xrightarrow{\tau}}$ and $P_j \xrightarrow{\mathtt{e}}$.

It is assumed that a type system is given to check that a process $P$ implements 
the contract $\sigma$. This is expressed by the judgment $\vdash P : \sigma$.

\begin{definition1} \label{def:consistency} 
A type system is \emph{consistent} if, whenever $\vdash P:\sigma$, we have

\begin{enumerate}
	\item $P\xrightarrow{\tau} P'$ implies $\vdash P':\sigma'$ and $\sigma \sqsubseteq \sigma'$;
	\item $P \xrightarrow{\alpha} P'$ implies $\vdash P':\sigma'$, $\sigma\xrightarrow{\alpha}$, and $\sigma(\alpha)\sqsubseteq \sigma'$;
	\item $P$ diverges implies $\sigma\Downarrow \emptyset$;
  \item $P\centernot{\xrightarrow{\tau}}$ implies $\sigma\Downarrow R$ and 
  $R \subseteq \{\alpha\ |\ P\xrightarrow{\alpha}\}$.
\end{enumerate}
\end{definition1}

For consistent type systems, the following Lemma has been proved.

\begin{lemma}[Subject reduction] \label{lemma:subjectreduction}
If $\vdash P: \rho$ and $\vdash Q:\sigma$ and $\rho\dashv \sigma$ and $P\ppar Q \xrightarrow {} P' \ppar Q'$, then $\vdash P': \rho'$ and $\vdash Q':\sigma'$ and $\rho'\dashv \sigma'$.
\end{lemma}

It has been shown that consistent type systems are sound with respect to compliance, i.e., two processes
are guaranteed to be compliant if their types are compliant, as formally stated by the following result.

\begin{theorem}
\label{th-types-and-compliance} If $\vdash P : \rho$ and $\vdash Q:\sigma$ and $\rho\dashv \sigma$ then $P \dashv Q$.
\end{theorem}

\section{Abstraction for contracts}\label{sec:abstractions}

We start by introducing a general definition of the notion of slicing or abstraction of concrete
processes.  We consider the language of concrete processes enriched with an operator that 
transforms any action over a hidden channel into an 
internal action. The abstraction operator is defined as follows
\[ \mathcal{A}_V[P] \]
where $V\subseteq \mathcal{N}$ is the set of visible actions. 

The process $\mathcal{A}_V[P]$ is a slice of $P$ that behaves as $P$ everytime $P$ performs an
 action over a visible port, while it performs an internal action when the subject 
 of the action executed by $P$ is a hidden channel. 
Consequently, we assume that the labeled transition system for processes is extended
with the following two rules

\[  \anmathrule{P\xrightarrow{\alpha} P'\quad   \alpha\in V}
               {\mathcal{A}_V[P]\xrightarrow{\alpha} \mathcal{A}_V[P']}
\qquad
  \anmathrule{P\xrightarrow{\alpha} P'\quad   \alpha\not\in V}
             {\mathcal{A}_V[P]\xrightarrow{\tau} \mathcal{A}_V[P']}
\]

%We remark here that the abstraction $\mathcal{A}_V[P]$ can evolve by executing a hidden action but a client cannot use those actions to 
%synchronize with the service. 

In addition, we define the effect of applying abstraction $\mathcal{A}_V$ over a contract $\sigma$ that
hides all actions of $\sigma$ that are not in $V$.

\begin{definition1}[Contract abstraction] The abstraction $\mathcal{A}_V$ of a contract $\sigma$, written $\mathcal{A}_V(\sigma)$, is inductively defined as follows:

\[
  \begin{array}{r@{\ =\ }l@{\hspace{2cm}}l}
    \mathcal{A}_V(0) & 0 \\
    \mathcal{A}_V(\alpha.\sigma) & \alpha.\mathcal{A}_V(\sigma) & \mbox{if }\ \alpha\in V\\
    \mathcal{A}_V(\alpha.\sigma) & \mathcal{A}_V(\sigma) & \mbox{if }\ \alpha\notin V\\
    \mathcal{A}_V(\Sigma_{i\in I} \alpha_i.\sigma_i) & \Sigma_{j\in J} \alpha_j.\mathcal{A}_V(\sigma_j) \oplus
                                 \bigoplus_{k \in K} \mathcal{A}_V(\sigma_k) \\
                               \multicolumn{3}{r}{ \mbox{with }\ J= \{i \in I| \alpha_i \in V\}\ \mbox{and}\
                                  K= \{i \in I| \alpha_i \not\in V\}}\\
    \mathcal{A}_V(\bigoplus_{i\in I} \alpha_i.\sigma_i ) & \bigoplus_{i \in I} \mathcal{A}_V(\alpha_i.\sigma_i)    
  \end{array} 
\]
\end{definition1}

Previous rules state that applying abstraction to a contract is not just removing the 
hidden actions. In fact, the abstraction of a contract accounts for the fact that a concrete 
process may commit a choice when executing a hidden action. The most interesting rule is the one for 
external choices. Note that the abstraction for $\sigma = \Sigma_{i\in I} \alpha_i.\sigma_i$ corresponds to a
contract that internally chooses whether to execute an internal action, i.e., some $\alpha_k\not\in V$,
or to leave the client to select one of the available visible actions $\alpha_j\in V$.

\begin{example} Consider the following variant of the service that handles loan requests  
described in the Introduction. In this variant, the service asks a third-party service for a recommendation based on client historical records. The third-party service 
responds back by sending either a positive or 
a negative feedback. A contract describing the behavior of the concrete service can be 
written as follows.

\[
   \sigma = {\it request}.\overline{\it askadvice}.({\it negative}.\overline{\it refused} + {\it positive}.\overline{\it approved}) 
\]

The corresponding contract describing the interaction of the service with the client
will be

\[
   \mathcal{A}_{\{ {\it request,refused, approved}\}}(\sigma) = {\it request}.(\overline{\it refused} \oplus \overline{\it approved}) 
\]

This abstraction states clearly that the loan service accepts a client 
request and then decides internally whether to approve or to refuse it. The internal 
choice in the abstraction reflects the fact that a service may commit a choice when it 
interacts over a hidden channel (e.g., it commits to refuse the request when it receives a 
negative feedback from the third party).
\end{example}

The following three results state properties for contract abstraction that will be used for 
proving main results of the paper. The next proposition relates the ready sets of the 
abstraction $\mathcal{A}_V(\sigma)$ with the ready sets of $\sigma$.

%\begin{proposition} $\mathcal{A}_V(\sigma) \Downarrow R$ if and only if:
%\begin{itemize}
%  \item there exists $S$ such that $\rho\Downarrow S$ and $R= V\cap S$; or
%  \item $\sigma\xmapsto{\alpha}\sigma'$ with $a\not\in V$ and $\mathcal{A}_V(\sigma')\Downarrow R$.
%\end{itemize}
%\end{proposition}

%\begin{proof} The proof follows by straightforward structural induction on $\sigma$.
%\end{proof}

\begin{proposition} 
\label{seqofhiddenreductions}
  $\mathcal{A}_V(\sigma) \Downarrow S$ if and only if
$\sigma\xmapsto{\alpha_1}\sigma_1\ldots\xmapsto{\alpha_n}\sigma_n$, $\alpha_1,..,\alpha_n \not\in V$ and 
$\sigma_n\Downarrow S'$ with $S'\cap V = S$.
\end{proposition}

\begin{proof} $\Rightarrow$) The proof follows by straightforward structural induction on $\sigma$. 
$\Leftarrow$) By induction on the length of the derivation. Base case follows by case analysis on the 
structure of $\sigma$. Induction step follows by case analysis on the structure of $\sigma$ and inductive hypothesis.
%\qed
\end{proof}

The following proposition characterizes the continuation $\mathcal{A}_V(\sigma)(\alpha)$ of an abstraction.
 
\begin{proposition} 
\label{redabstraccion}
  $\mathcal{A}_V(\sigma) \xmapsto{\alpha}\rho$ if and only if
  $\alpha\in V$, $\rho = \bigoplus_{\rho_i\in \mathtt{Alc}(\sigma,\alpha,V)} \mathcal{A}_V(\rho_i)$ with
 \[\mathtt{Alc}(\sigma,\alpha,V) = \{\sigma'\ |\ \sigma\xmapsto{\beta_1}\sigma_1\ldots\xmapsto{\beta_n}\sigma_n\xmapsto{\alpha}\sigma'\ \mbox{and}\ \beta_1,\ldots,\beta_n\not\in V\}  \neq\emptyset.\]
\end{proposition}

\begin{proof} $\Rightarrow$) The proof follows by straightforward structural induction on $\sigma$. 
$\Leftarrow$) By induction on the length of the derivation.  Base case follows by case analysis on the 
structure of $\sigma$. Induction step follows by case analysis on the structure of $\sigma$ and inductive hypothesis. 
%\qed
\end{proof}

The result below shows that abstraction preserves continuations under visible actions.

\begin{proposition} 
\label{prop:abstofvisiblered}
Let $\sigma\xmapsto{\alpha}$ and $\alpha\in V$. Then, $\mathcal{A}_V(\sigma) \xmapsto{\alpha} \mathcal{A}_V(\sigma(\alpha))$.
\end{proposition}

\begin{proof} The proof follows by straightforward structural induction on $\sigma$.
%\qed
\end{proof}

The following proposition ensures that abstraction preserves subcontract relation or, in 
other words, states that if one contract can be safely replaced by another contract, 
then any possible slice of the original contract can be safely replaced by the corresponding
slice of the new contract. 

\begin{proposition} 
\label{prop:abstractionsubcontract}
If $\sigma \sqsubseteq \rho$ then  $\mathcal{A}_V(\sigma) \sqsubseteq \mathcal{A}_V(\rho)$.
\end{proposition}

\begin{proof}
The proof follows by showing that $\mathcal{S}= \{(\mathcal{A}_V(\sigma),\mathcal{A}_V(\rho)) | \sigma \sqsubseteq \rho\}$ is a subcontract relation. 
Due to space limitation we omit details here. (We report proof in Appendix~\ref{app:proofs}).
%\qed
\end{proof}

The following two propositions state properties about the continuations of contract abstractions. These two results
are used in the proof of the main result of the following section (Proof details can be found in Appendix \ref{app:proofs}). 

\begin{proposition} 
\label{prop:abstractionsubcontracthiddenreduction}
If $\sigma(\alpha) \sqsubseteq \rho$ and $\alpha\not\in V$ then  $\mathcal{A}_V(\sigma) \sqsubseteq \mathcal{A}_V(\rho)$.
\end{proposition}

\begin{proposition} 
\label{prop:abstractionsubcontractvisiblereduction}
If $\sigma(\alpha) \sqsubseteq \rho$ and $\alpha\in V$ then  $\mathcal{A}_V(\sigma)(\alpha) \sqsubseteq \mathcal{A}_V(\rho)$
\end{proposition}

%\begin{proposition} If $\sigma(\alpha) \sqsubseteq \sigma'$  and $\alpha \in V$, then  $\mathcal{A}_V(\sigma(\alpha))
%\sqsubseteq \mathcal{A}_V(\sigma')$.
%\end{proposition}

Finally, we show how to extend a consistent type system in order to be able to type processes that use abstraction. 
This is achieved by extending any consistent type system for concrete processes with the following typing rule 
\[
   \mathrule{TypeAbstraction}{\vdash P : \sigma}{\vdash\mathcal{A}_V[P] : \mathcal{A}_V(\sigma)}
\]

Next result shows that the above rule preserves consistency. 

\begin{proposition} \label{prop:enrichedconsistency} A consistent type system enriched with rule {\sc(TypeAbstraction)}
results in another consistent type system.
\end{proposition}

\begin{proof}
   Let $\vdash P:\sigma$. As regards consistency condition (1), assume $\mathcal{A}_V[P]\tr{\tau} \mathcal{A}_V[P']$, then either $P\tr{\tau} P'$ or
$P \tr{\alpha} P'$ with $\alpha\not\in V$. When $P\tr{\tau} P'$, consistency
ensures that $\vdash P :\sigma'$ and  $\sigma \sqsubseteq\sigma'$. By Proposition~\ref{prop:abstractionsubcontract}, $\mathcal{A}_V(\sigma)\sqsubseteq \mathcal{A}_V(\sigma')$. If $P \tr{\alpha} P'$ then by consistency $\vdash P :\sigma'$ and  $\sigma(\alpha) \sqsubseteq\sigma'$. By Proposition~\ref{prop:abstractionsubcontracthiddenreduction}, $\mathcal{A}_V(\sigma)\sqsubseteq \mathcal{A}_V(\sigma')$.   As regards
consistency condition (2), assume that $P\tr{\alpha} P'$ and $\alpha\not\in V$. By consistency, 
$\vdash P :\sigma'$, $\sigma(\alpha)\sqsubseteq\sigma'$. By Proposition~\ref{prop:abstractionsubcontractvisiblereduction}, 
$\mathcal{A}_V(\sigma)(\alpha)\sqsubseteq\mathcal{A}_V(\sigma')$. As regards consistency
condition (3), assume that $\mathcal{A}_V[P]$ diverges. Then either $P$ diverges or $P$ has an infinite 
derivation $P\tr{\alpha_1} P_1 \ldots \tr{\alpha_n} P_n \tr{\alpha_{n+1}} \ldots$ with $\alpha_i = \tau$ or $\alpha_i\not\in V$.
If $P$ diverges, then $P\Downarrow \emptyset$. Therefore, $\mathcal{A}_V[P]\Downarrow \emptyset$. 
Otherwise, assume $P$ has an infinite 
derivation $P\tr{\alpha_1} P_1 \ldots \tr{\alpha_n} P_n \tr{\alpha_{n+1}} \ldots$ with $\alpha_i = \tau$ or $\alpha_i\not\in V$. 
By consistency, this implies that there exists an infinite derivation for the contract $\sigma\xmapsto{\alpha_1}\sigma_1\ldots\xmapsto{\alpha_n}P_n\xmapsto{\alpha_{n+1}}\ldots$ but this is not possible, since we are considering finite contracts.  Finally, as regards consistency condition (4),
assume that $\mathcal{A}_V[P]\nt{\tau}$. Then,  $P\nt{\tau}$ and $P\nt{\alpha}$  for all $\alpha\not\in V$.  We derive $\sigma\Downarrow R$ 
where $R \subseteq \{\alpha \ | \ P \tr{\alpha}\}$. Moreover $R \subseteq V$ since $P\nt{\alpha}$  for all $\alpha\not\in V$. 
By proposition~\ref{seqofhiddenreductions}, $\mathcal{A}_V[P]\Downarrow R$. Since, $\alpha\in R$ implies $\alpha\in V$, $P\tr{\alpha}$
implies $\mathcal{A}_V[P]\tr{\alpha}$. Hence, $R \subseteq \{\alpha \ | \ \mathcal{A}_V[P] \tr{\alpha}\}$.
%\qed
\end{proof}

%A questo punto abbiamo come corollario che

%If $\vdash P:rho$, $\vdash Q:\sigma$ and $\rho \dashv A_V(sigma)$ then
%$P \dashv A_V(Q)$

%A questo punto ci manca far vedere che

%$A <^V P$ and $\vdash P:\sigma$ then $\vdash A: rho$ with $\rho
%\sqsubseteq A_V(sigma)$.

%questo farebbe il link fra i due (non ho pensato a fare la prova).

\section{Contracts for abstract processes}\label{sec:typesystem}

In this section we aim at bridging the theories of processes and contracts presented in the previous sections. 
 We remark that although the language of abstract processes is a kind of value-passing {\sc  ccs},  the remaining of this section will consider just finite domains for values, and hence we implicitly will refer to  the usual encoding of value-passing {\sc  ccs} into {\sc  ccs} (i.e., we will refer a  channel and a tuple of values  just as a single action).  Moreover, we say an action is a visible action if its subject is a visible name.

We define a type system that assigns contracts to processes and we prove that the proposed type system is consistent according to Definition~\ref{def:consistency}.
We use judgments of the form $\vdash P: \sigma$. We report the typing rules in Table \ref{table:type_system} (Rules are analogous to the type system for {\sc ws-bpel} proposed in~\cite{CGP09:TCWS}).   
The main idea behind the type system is that types can contain neither $\tau$'s nor parallel composition, and that the type of a guarded choice must be an internal choice if its guards are $\tau$'s. In this 
sense, rule ({\sc tau}) is as expected. On the other side, rule ({\sc pref}) allows recording in the contract any non-$\tau$ prefix. Rules ({\sc sum}) and ({\sc par}) are the most interesting and account for assigning to both external choice and parallel composition a contract that is a suitable internal choice. Specifically, the type of a choice is 
obtained as an internal choice between the branches with $\tau$'s as prefixes and an external choice of visible prefixed branches. 
For instance, consider the process $P \equiv a.P_1 + b.P_2 + \tau.P_3$. It holds that $\vdash P: (a.\sigma_1 + b.\sigma_2) \oplus \sigma_3$ for $ \vdash P_1:\sigma_1$, $\vdash P_2:\sigma_2$, and $\vdash P_3:\sigma_3$.
% but if we exclude $b$ from the visible names, we have that $ \{a,c \}, \true \vdash P: (a.\sigma_1 + c.\sigma_3) \oplus b.\sigma_2$. 
Rule ({\sc par}) exploits an idea that reminds the expansion lemma, namely the executions performed by a parallel composition $P|Q$ are the sum of the executions of $P_i \,|\,Q$ and $P \,|\,Q_j$, being $P_i$ and $Q_j$ all the continuations of $P$ and $Q$, respectively. Note that we do not consider the executions resulting from synchronizations of $P$ and $Q$ over complementary actions, as such synchronizations within the same orchestrator are not allowed. Akin to rule ({\sc sum}), the type of a parallel composition is the external choice of the non-$\tau$ prefixed alternatives and the internal choice of the branches whose prefixes are $\tau$'s. Note that rule ({\sc par}) requires to consider all possible computations 
$P \ltr{\act_i} P_i$ and $Q \ltr{\beta_j} Q_j$ and, consequently, this rule is well-defined when we have a finite number of such computations. We remark that the language for concrete and abstract processes that we are considering ensures
us that all processes are finitely branching, hence rule ({\sc par}) is well-defined for our target language. 

As an example, $\vdash (a + \tau) \, | \, (b + c) : (a.\sigma_{b+c} + b.\sigma_{a + \tau} + c.\sigma_{a + \tau}) \oplus \sigma_{b+c}$. 
Rules ({\sc cond1}), ({\sc cond2}), and ({\sc cond3}) concern the type of conditional statements. 
More in detail, rule ({\sc cond1}) applies if $\opaque\in\{m, n \}$. In this case, the type of the if-then-else is the internal choice between the type of the two possible alternatives. Conversely, rules ({\sc cond2}) and ({\sc cond3}) state that $m$ and $n$ are both visible, then the type assigned is the type of the only possible branch.

\begin{table}[t]
\begin{math}
\begin{array}{l}
 \mathaxiom{Nil}{\vdash {0}: 0}\\
  \\
  \mathrule{Tau}
  {\vdash P : \sigma}
  {\vdash \tau.P : \sigma}\\
  \\
  \mathrule{Pref}
  {\vdash P : \sigma \qquad \act \neq \tau}
  {\vdash \act.P : \act.\sigma}\\
  \\
  \mathrule{Sum}
  {\act_i \neq\tau \qquad \vdash P_i:\sigma_i \qquad \vdash Q_j:\rho_j}
  {\vdash \Sigma_{i\in I} \act_i.P_i + \Sigma_{j \in J} \tau.Q_j: \Sigma_{i \in I} \act_i.\sigma_i \oplus 
   \bigoplus_{j\in J} \rho_j}\\ 
  \\
  \mathrule{Par}
  {\vdash P_i | Q:\sigma_i\ {\it for\  all\ } P \ltr{\act_i} P_i   \qquad \vdash P | Q_j:\rho_j\     
  {\it for\  all\ }  Q \ltr{\beta_j} Q_j}
  {\vdash P | Q : (\Sigma_{\act_i \neq \tau} \act_i.\sigma_i + \Sigma_{\beta_j \neq\tau} 
   \beta_j.\rho_j)
   \oplus  \bigoplus_{\act_i = \tau} \sigma_i \oplus \bigoplus_{\beta_j = \tau} \rho_j }
  \qquad
  {\mbox{where:}\Bigg\{ 
    \begin{array}{l}
      P \ltr{\act_i} P_i\\
      Q \ltr{\beta_j} Q_j
    \end{array}}\\
   \\ 
   \mathrule{cond1}
  {\vdash P : \sigma \qquad \vdash Q : \rho \qquad \opaque\in\{m,n\}}
  {\vdash\ifte{m=n}PQ : \sigma \oplus \rho}\\
  \\
  \mathrule{cond2}
  {\vdash P : \sigma \qquad \qquad \opaque\not\in\{m,n\} \qquad \qquad m = n}
  {\vdash\ifte{m=n}PQ : \sigma}\\
  \\
  \mathrule{cond3}
  {\vdash Q : \rho \qquad \qquad \opaque\not\in\{m,n\} \qquad \qquad m \neq n}
  {\vdash\ifte{m=n}PQ : \rho}\\
  \\  
 \end{array}
\end{math}
\caption{Type System for Contracts}
\label{table:type_system}
\end{table}

\begin{theorem}\label{thm:consistency}
 The type system $\vdash P:\sigma$ shown in Table \ref{table:type_system} is consistent. 
\end{theorem}

\begin{proof} The proof is by induction on the structure of $P$. See appendix \ref{app:proofs}.  
%\qed
\end{proof}

Next result states an auxiliary property that will be used when proving the main result of this section. It states that the reductions of an abstraction of a concrete process are in one-to-one correspondence with the visible reductions of the concrete process. 

\begin{proposition} \label{prop:abstractionandreductions} 
 Let $P$ and $Q$ be two closed processes such that $P\propto^V Q$. 
 \begin{enumerate}
  \item $\mathcal{A}_V[Q] \tr{\alpha} \mathcal{A}_V[Q']$ implies $P\tr{\alpha} P'$ and $P'\propto^V Q'$.
  \item $P\tr{\alpha} P'$ implies $\mathcal{A}_V[Q] \tr{\alpha} \mathcal{A}_V[Q']$ and $P'\propto^V Q'$.
 \end{enumerate} 
\end{proposition}

\begin{proof} 
 See Appendix \ref{app:proofs}.
%\qed
\end{proof}

The following result formalizes the relation among abstractions and strong compliance. It basically states that whenever a client $P$ has a type that is compliant with the type of an abstract process  $Q$ which is an abstraction of a concrete process $R$, then $P$ correctly interacts with the filtered process  $\mathcal{A}_V[R]$

\begin{theorem} 
 Let $P:\sigma$,  $Q:\rho$ and $Q \propto^V R$. If $\sigma\dashv \rho$ then $P \dashv \mathcal{A}_V[R]$. 
\end{theorem}

\begin{proof} The proof follows the line of the proof of Theorem 4.5 in~\cite{CGP09:TCWS}.
 Akin to ~\cite{CGP09:TCWS}, we reserve a special action $e$ (for ``end'') that can occur in 
 client contracts and that represents the ability of the client to successfully terminate. 
 Then we require that, whenever no further interaction is possible between the client and 
 the service, the client be in a state where this action is available.

 First, we notice that, by Proposition~\ref{prop:abstractionandreductions}, any computation 
 $P || \mathcal{A}_V[R] \rightarrow P' || \mathcal{A}_V[R']$ has a corresponding computation
 $P || Q \rightarrow P' || Q'$ with $Q' \propto^V R'$ .
% Consequently, any maximal computation of $P || A_V[R]$ is a maximal computation of $P || Q$ with $Q \propto^V R$.
 Because of Lemma~\ref{lemma:subjectreduction}, we only need to 
 consider maximal computations, i.e., cases in which $P\ppar \mathcal{A}_V[R] \centernot{\xrightarrow{}}$ or $P\ppar \mathcal{A}_V[R]$ diverges 
 (equivalently, cases in which $P\ppar Q \centernot{\xrightarrow{}}$ or $P\ppar Q$ diverges for $Q \propto^V R$).
 Let $P\ppar Q \centernot{\xrightarrow{}}$ and assume, 
by contradiction, that $P \centernot{\xrightarrow{\mathtt{e}}}$. From $\sigma \dashv \rho$ we know that $\rho\downarrow R$ 
implies $R\neq \emptyset$ (by Definition~\ref{def:strongcompliance}). From 
$P\ppar Q\centernot{\xrightarrow{}}$ , we have that whenever $P\xrightarrow{\alpha}$ we have $Q\centernot{\xrightarrow{\alpha}}$ and hence $\mathcal{A}_V[R]\centernot{\xrightarrow{\alpha}}$. Consequently, $\{\alpha | P \xrightarrow{\alpha}\} \cap \mathtt{co}(\{\alpha | Q \xrightarrow{\alpha}\}) = \emptyset$. 
From consistency condition (4) there exist $R$ and $S$ such that 
$\rho\Downarrow R$ and $\sigma\Downarrow S$ and $\mathtt{co}(R)\cap S = \emptyset$
and $\mathtt{e}\not\in R$, but this is absurd from the hypothesis that $\rho\dashv \sigma$. Hence $P\xrightarrow{\mathtt{e}}$.
Assume $P\ppar Q$ diverges. First, note that $P$ cannot diverge since consistency 
condition (3) requires that $\rho\Downarrow\emptyset$.  Then, the only possibility is $P\centernot{\xrightarrow{}}$ 
and $Q$ diverges. By consistency condition (3) we derive $\sigma\Downarrow \emptyset$,
hence $\rho\Downarrow R$ implies $\mathtt{e}\in R$. From consistency condition (4)
we conclude $P\xrightarrow{\mathtt{e}}$. 
%\qed
\end{proof}

\section{Conclusions}
In this paper we have investigated the relation among the theory of contracts and the hiding of selected actions.  We have shown that we can recover the notion of abstraction as a kind of filter over processes and we accommodate this notion into the theory of contracts for web services when considering finite contracts. We remark that the current definition for abstraction is not suitable for handling infinite contracts. In fact, it turns out that abstraction may not preserve the contractivity condition of contracts. In order to see this, consider the contract $\sigma = b+ a.b + a.a.b + a.a.a.b + \ldots+ a.a.a\ldots $ that accounts for an infinite execution of  $a$'s. Contract $\sigma$ can be written with the recursive expression $\mathit{rec}\  x = a. x + b$. Then, by taking the current definition of abstraction, $\mathcal{A}_{\{b\}} (\sigma)$ will be associated with the recursive equation $\mathit{rec}\  x = x+b$, for which contractivity does not hold.  We left as future work the definition of abstraction for infinite contracts. 

\paragraph{Acknowledgements} The authors thank
anonymous reviewers for their helpful comments on an earlier version of this paper.

\bibliographystyle{eptcs} % or whatever you prefer \bibliographystyle{abbrv}

\appendix

\section{The non-symbolic semantics of orchestrators} \label{app:non-symb}

The original definition of the operational semantics of orchestrators as defined in~\cite{BM:APOL}
is shown in Figure~\ref{figure:LTSCP}

 \begin{figure}[tp]
 \[
 \begin{array}{l}
   \mathaxiom{tau}
   {\tau.P \tra {\tau} P}
 \quad
  \mathaxiom{out}
  {\outp{x}{\tuple{a}}.P \tra {\outp{x}{\tuple{a}}} P}
 \quad
  \mathaxiom{in}
  {\inp{x_1}{\tuple{v_1}}.P_1+ \ldots+\inp{x_n}{\tuple{v_n}}.P_n
   \tra {\finp{x_i}{\tuple{a}}} P_i\{\tuple{a}/\tuple{v_i}\}}
 \\[5pt]
   \mathrule{if}
         {P\tra {\alpha} P' }
         {\ifte{a=a}{P}{Q} \tra {\alpha} P'}
 \hfill
   \mathrule{else}
   	    {Q\tra {\alpha} Q' \quad a\neq b }
         {\ifte{a=b}{P}{Q} \tra {\alpha} Q'}
 \\[5pt]
 \mathrule{par}
 				 {P\tra {\alpha} P'}{P\;|\;Q\tra {\alpha}P'\;|\;Q}
 \hfill
 \mathrule{str}
 	{P\equiv Q\quad Q\tra{\alpha} Q'\quad Q'\equiv P'}
         {P\tra{\alpha} P'}
\\[5pt]
  \mathruletwo{choice-1}
  {P \tra {\alpha} P' \quad  \opaque\in\{m,n\}}
  {\ifte{m=n}{P}{Q} \tra { \alpha} P'}
  \hfill 
  \mathruletwo{choice-2}
  {Q \tra {\alpha} Q' \quad  \opaque\in\{m,n\}}
  {\ifte{m=n}{P}{Q} \tra {\alpha} Q'}
  \\[10pt]
  \mathaxiom{choice-3}
  {\inp{x_1}{\tuple{v_1}}.P_1+
  \ldots+\inp{\opaque}{\tuple{v_i}}.P_i+\ldots+\inp{x_n}{\tuple{v_n}}.P_n
  \tra {\tau} P_i}
 \end{array}
 \]
 \caption{LTS for processes}
 \label{figure:LTSCP}
 \end{figure}
 
 Following result states the correspondence between the original semantics non-symbolic semantics
 and the one introduced in Definition~\ref{def:alt-non-symb-sem}.

\begin{theorem} Let $P$ be a closed process. $P\tr {\alpha} P'$ iff $P\tra {\alpha} P'$.
\end{theorem}

\begin{proof}

  \begin{itemize} 
    \item[$\Rightarrow$)] By Definition~\ref{def:alt-non-symb-sem}, $P\tr {\alpha} P'$ implies $P\tr {M, \lambda} Q$ and  $\alpha = \lambda\sigma$ and 
    $P'=Q\sigma$ with $\sigma \models M$. The proof follows by straightforward rule induction on the derivaion of $P\tr {M, \lambda} Q$.
       \begin{itemize}
          \item {\bf({\sc s-tau})}: $P = \tau.Q$, $M= true$, $\lambda=\tau$. For any substitution $\sigma$, we have that 
          $\sigma \models M$, $\alpha = \lambda\sigma = \true$. Also, $P$ closed implies $Q$ closed, hence $P' = Q \sigma = Q$. By rule
          ({\sc tau}), $P = \tau.Q \tra{\tau} Q$. 
          \item {\bf({\sc s-out}) and \bf({\sc s-choice-3})}: these cases follow as for ({\sc s-tau}).
          \item {\bf({\sc s-in})}: $P={\inp{x_1}{\tuple{v_1}}.P_1+ \ldots+\inp{x_n}{\tuple{v_n}}.P_n}$, $M=true$, $\lambda=\inp{x_i}{\tuple{v_i}}$,
          $Q = P_i$. Since $P$ is closed, $fn(Q) \subseteq \tuple{v_i}$. Then, for any $\sigma \models M$, $P'= Q\sigma = P_i\sigma = P_i\sigma_{|\tuple{v_i}}$ and $\alpha = \lambda\sigma = \lambda\sigma_{|\tuple{v_i}}$. By rule ({\sc in}) $P \tra {\inp{x_i}{\tuple{v_i}}\sigma_{|\tuple{v_i}}} P_i\sigma_{|\tuple{v_i}} = P'$
          \item {\bf({\sc s-par})}: $P = P_1 | P_2$, $P_1\tr{M,\lambda} P_1'$,
          $Q = P_1'|P_2$. Since, $\sigma \models M$, by Definition~\ref{def:alt-non-symb-sem} $P_1\tr{\alpha} P_1'\sigma$. 
          By inductive hypothesis,  $P_1\tr{\alpha} P_1'\sigma$. By rule ({\sc par}) $P_1|P_2\tra{\alpha} P_1'\sigma|P_2$. Since
          $P$ is closed, also $P_2$ is closed. Hence, $P_2\sigma = P_2$. Therefore, $P_1|P_2\tra{\alpha} Q\sigma$
          \item {\bf({\sc s-str}), \bf({\sc s-choice-1}) and \bf({\sc s-choice-2})}: these cases follow analogously to ({\sc s-par}).
          \item {\bf({\sc s-if})}: Since $P$ is closed, the only possibility for $m$ and $n$ is to be the same constant. Hence, $P = \ifte{a=a}{P_1}{Q_2}$, $P_1\tr{M,\lambda} Q$. By inductive hypothesis, $P_1\tra{\alpha} Q\sigma$. Then, by rule ({\sc if}),  $P\tra{\alpha} Q\sigma$.
          \item {\bf({\sc s-else})}: This case is analogous to ({\sc s-if}).
       \end{itemize}

    \item[$\Leftarrow$)] 
       \begin{itemize}
          \item {\bf({\sc tau})}: $P= \tau.P'$, $\alpha = \tau$. By ({\sc tau}), $P\tr{true, \tau} P'$. Since, $P'$ is closed, $P'\sigma = P'$ for any $\sigma$.
          $P\tr{\tau} P'$.
          \item {\bf({\sc out})}: This case follows as ({\sc tau}).
          \item {\bf({\sc in})}: $P={\inp{x_1}{\tuple{v_1}}.P_1+ \ldots+\inp{x_n}{\tuple{v_n}}.P_n}$, $\alpha = \outp{x_i}{\tuple{a}}$ and $P'= P_i \{\tuple{a}/\tuple{v_i}\}$.By rule ({\sc in}), $P\tr{true,{\inp{x_1}{\tuple{v_1}}}} P_i$. Note that $\{\tuple{a}/\tuple{v_i}\} \models true$. Then, by Definition~\ref{def:alt-non-symb-sem}, $P\tr{\alpha} P_i\{\tuple{a}/\tuple{v_i}\}$.
          \item {\bf({\sc if})}: $P = \ifte{a=a} {P_1}{P_2}$, $P_1\tra{\alpha} P'$. By
          inductive hypothesis, $P_1\tr{\alpha} P'$. By definition, there exist 
          $M$, $Q$, $\sigma$ and $\lambda$ s.t. $\sigma\models M$, $\alpha = \lambda\sigma$, $P' = Q\sigma$, $P_1 \tr{M, \lambda} Q$. Since $M$ is consistent,
          $M\wedge a=a$ is consistent. Then, by rule ({\sc s-in}), $P = \tr{M\wedge a=a, \lambda} Q$. From $\sigma \models M$, we have $\sigma \models M\wedge a=a$. By Definition~\ref{def:alt-non-symb-sem}, $P\tr{\alpha} P'$.  
          \item {\bf({\sc else})}:Analogous to ({\sc if}).
          \item {\bf({\sc par})}: $P =P_1|P_2$ with $P_1\tra{\alpha} P_1'$ and $P' = P_1'|P_2$. By inductive hypothesis, $P_1\tr{\alpha} P_1'$.
          By definition~\ref{def:alt-non-symb-sem}, $P_1\tr{M,\lambda} Q$ and there exists $\sigma\models M$ s.t. $\alpha = \lambda\sigma$ and 
          $P_1 = Q\sigma$. By rule ({\sc s-par}), $P_1|P_2\tr{M,\lambda} Q|P_2$ (side condition holds because $P_2$ is closed). By definition~\ref{def:alt-non-symb-sem}, we have $P_1|P_2\tr{alpha} (Q|P_2)\sigma$. Since $P_2$ is closed. $P_2\sigma = P_2$ and, hence, 
          $(Q|P_2)\sigma = P'$
          \item {\bf({\sc str}),\bf({\sc choice-1}),\bf({\sc choice-2}) and \bf({\sc choice-3})}: Follows by using inductive hypothesis. 
       \end{itemize}

  \end{itemize}
%\qed
\end{proof}

\begin{proposition} 
\label{prop:fninreductions}If $P\tr{M,\lambda} Q$ then 
\begin{itemize}
   \item $\fn(M) \subseteq \fn(P)$.
   \item $\fn(Q) \subseteq \fn(P)\cup\bn(\lambda)$.
   \item $M$ is consistent.

\end{itemize}
\end{proposition}

\begin{proof} It follows by straightforward rule induction.
%\qed
\end{proof}

\begin{proposition} 
\label{prop:condition-cut}
If $P\propto^V_M Q$ and $\fn(P,Q) \cap \fn(M) = \emptyset$ then
$P \propto^V Q$.
\end{proposition}

\begin{proof}  We first fix the following notation: given a constraint $M$ and a set of 
names $S$, we write $M\backslash S$ from $M$ by removing all terms containing a name in $S$.
If $P\propto^V_M Q$ then there exists a family of abstraction relations 
$\{\mathcal{R}^V_N\}_N$ such that $P \mathcal{R}^V_M Q$.
We take the following family of relations $\{\mathcal{S}^V_L\}_L$, where 
$\mathcal{S}^V_L = \mathcal{R}^V_N$ with $L= N\backslash\fn(M)$. We now show that this is 
a family of abstractions relations. 

Let $P$ and $Q$ such that $P\mathcal{S}^V_L Q$:

 \begin{enumerate}
 \item Assume $Q \tr {N_1, \act} Q'$ and $\bn(\act)\cap\fn(P,Q,L) =\emptyset$. Without loss of 
 generality we can
 assume that $\bn(\act)\cap\fn(P,Q,L\cup M) = \emptyset$ (This can always be  
 achieved by alpha-renaming bound names.). We know that $P\mathcal{R}^V_N Q$ with
 $L = N\backslash\fn(M)$. Since $\mathcal{R}^V_N$ is an abstraction relation, 
 there exists a $N\wedge N_1$-decomposition $D$ s.t. $\forall M_1\in D$ there exists $P \tr {N_1',\act'} P'$, with
   $M_1\Rightarrow N_1'$, $\act_{|V} = \act'$ 
   and $P' \mathcal{R}^{V \cup \vn(\act) }_{\;M_1} Q'$. By Proposition~\ref{prop:fninreductions},
   $P \tr {N_1',\act'} P'$ implies $\fn(N_1') \subseteq \fn(P)$. Since, $\fn(P)\cap\fn(M) = \emptyset$ then $\fn(N_1')\cap \fn(M) = \emptyset$ for all $N_1'$. Consequently, $M_1\backslash\fn(M) \Rightarrow N_1'$ and $\bigvee_i M_i\backslash\fn(M)$ is a $L$-decomposition.
   
 \item if $P \tr {N, \act} P'$, the proof follows as in the previous case. 
\end{enumerate}
%\qed
\end{proof}

\begin{proposition} 
\label{prop:symbinstantiation}
If $P\{a/x\}\tr{M,\lambda} Q$ then there exist $N$, $\lambda'$ and $Q'$ s.t.
$M=N\{a/x\}$, $\lambda = \lambda'\{a/x\}$ and $Q= Q'\{a/x\}$. 
\end{proposition}

\begin{proof} The proof follows by straightforward rule induction.
%\qed
\end{proof}

\begin{proposition} 
\label{prop:visible-cut}
Let $P\propto^{V\cup\{x\}}_M Q$. For all $a$ s.t. $\{a/x\}\models M$, 
 $P\{a/x\}\propto^{V}_M Q\{a/x\}$.
\end{proposition}

\begin{proof}  We take the following family of relations $\{\mathcal{S}^V_L\}_L$, where 

\[\mathcal{S}^V_M = \{(P\{a/x\},Q\{a/x\}) | P \mathcal{R}^{V\cup\{x\}}_M Q\}\]
We show that this is 
a family of abstractions relations. 
Assume that $P\{a/x\}\mathcal{S}^V_M Q\{a/x\}$:

 \begin{enumerate}
 \item Assume $Q\{a/x\} \tr {N_1, \act} Q'$ and $\bn(\act)\cap\fn(P,Q,N) =\emptyset$. By Proposition~\ref{prop:symbinstantiation},
 $Q \tr {N_0, \act_0} Q_0$ and $N_1 = N_0\{a/x\}$, $\act = \act_0\{a/x\}$ and $Q'=Q_0\{a/x\}$. Since
  $P \mathcal{R}^{V\cup\{x\}}_M Q$, there is a $M\wedge N_0$-decomposition $D$ s.t. $\forall M'\in D$
  there exists $P\tr{N_0',\act_0'} P_0'$ with $M'\Rightarrow N_0'$, $\act_{|V} = \act'$ and
   $P_0' \mathcal{R}^{V\cup\{x\}\cup\vn(\act)}_{M'} Q_0$. By definition, $P_0\{a/x\}\mathcal{S}^{V\cup\vn(\act)\backslash\{x\}}_{M'} Q\{a/x\}$,
   From  $M' \Rightarrow N_0$ we have that $M'\{a/x\} \Rightarrow N_0\{a/x\}$. Since $D$ is a $M\wedge N_0$-decomposition
   we have $M\wedge N_0 \Rightarrow D$ and hence $(M\wedge N_0) \{a/x\}\Rightarrow D\{a/x\}$. From
   $\{a/x\}\models M$ we have that $M\wedge N_0 \{a/x\}\Rightarrow D\{a/x\}$. Consequently, 
   $D\{a/x\}$ is the requested $M\wedge N_1$-decomposition. 
 \item if $P\{a/x\} \tr {N, \act} P'\{a/x\}$, the proof follows as in the previous case. 
\end{enumerate}
%\qed
\end{proof}

\section{Proofs of the results in Sections \ref{sec:abstractions} and \ref{sec:typesystem}} \label{app:proofs}

\paragraph{Proof of Proposition \ref{prop:abstractionsubcontract}.}
The proof follows by showing that $\mathcal{S}= \{(\mathcal{A}_V(\sigma),\mathcal{A}_V(\rho)) | \sigma \sqsubseteq \rho\}$ is a subcontract relation. 
\begin{enumerate}
 \item Assume $\mathcal{A}_V(\rho) \Downarrow R$. By Proposition~\ref{seqofhiddenreductions}, we have that
 $\rho\xmapsto{\alpha_1}\rho_1\ldots\xmapsto{\alpha_n}\rho_n$ and $\rho_n\Downarrow R'$ with $R'\cap V = R$.
 Since  $\sigma \sqsubseteq \rho$, there exists 
 $\sigma\xmapsto{\alpha_1}\sigma_1\ldots\xmapsto{\alpha_n}\sigma_n$ with $\sigma_i \sqsubseteq \rho_i$ for $i= 1..n$.
 hence, there exists $S'\subseteq R'$ such that $\sigma_n\Downarrow S'$. By Proposition~\ref{seqofhiddenreductions},
 $\mathcal{A}_V(\sigma) \Downarrow S$ with $S=S'\cap V$. Since $S'\subseteq R'$ we have that $S = S'\cap V \subseteq R'\cap V = R$.
 \item Assume $\rho\xmapsto{\alpha}\rho'$. By Proposition~\ref{redabstraccion},   
  $\alpha\in V$ and $\rho' = \bigoplus_{\rho_i\in \mathtt{Alc}(\rho,\alpha,V)} \mathcal{A}_V(\rho_i)$ with
 \[\mathtt{Alc}(\rho,\alpha,V) = \{\rho'\ |\ \rho\xmapsto{\beta_1}\rho_1\ldots\xmapsto{\beta_n}\rho_n\xmapsto{\alpha}\rho'\ \mbox{and}\ \beta_1,\ldots,\beta_n\not\in V\}.\]
 Since $\sigma \sqsubseteq \rho$, for any $\rho_i$ in $\mathtt{Alc}(\rho,\alpha,V)$ there exists 
 a $\sigma_i$ s.t. $\sigma_i\in\mathtt{Alc}(\sigma,\alpha,V)$, i.e., $\sigma\xmapsto{\beta_1}\tau_1\ldots\xmapsto{\beta_n}\tau_n\xmapsto{\alpha}\sigma_i$.
 By proposition~\ref{redabstraccion},
 $\sigma'= \bigoplus_{\sigma_i\in\mathtt{Alc}(\sigma,\alpha,V)} \mathcal{A}_V(\sigma_i)$.
 It remains to show that $(\sigma', \rho')\in \mathcal{S}$. This is done by noting that
 $\sigma' = \bigoplus_j\mathcal{A}_V(\tau_j) \oplus  \tau$ such that any $\tau_j \in \mathtt{Alc}(\sigma,\alpha,V)$ 
 and there is a corresponding $\rho_j$ in $\mathtt{Alc}(\rho,\alpha,V)$ and $\tau_j\sqsubseteq \rho_j$. Note that it can be easily proved that $\sigma_1 \sqsubseteq \rho_1$ and $\sigma_2 \sqsubseteq \rho_2$ implies $\sigma_1\oplus\sigma_2 \sqsubseteq\rho_1\oplus \rho_2$. Consequently,
 $\bigoplus_j \tau_j \sqsubseteq \bigoplus_j\rho_j$. We can easily also prove that $\sigma_1\oplus \sigma_2\sqsubseteq \sigma_1$ for all 
 $\sigma_1, \sigma_2$. Hence, $\bigoplus_j \tau_j \oplus \tau \sqsubseteq \bigoplus_j\rho_j$, and finally, $(\sigma',\rho')\in\mathcal{S}$ by definition of $\mathcal{S}$
\end{enumerate}

\paragraph{Proof of Proposition \ref{prop:abstractionsubcontracthiddenreduction}.}
The proof follows by showing that $\mathcal{S}= \{(\mathcal{A}_V(\sigma),\mathcal{A}_V(\rho)) | \sigma(\gamma) \sqsubseteq \rho \ \mbox{and}\ \alpha\in V\}$ 
is a subcontract relation. 
\begin{enumerate}
 \item Assume $\mathcal{A}_V(\rho) \Downarrow R$. By Proposition~\ref{seqofhiddenreductions}, we have that
 $\rho\xmapsto{\alpha_1}\rho_1\ldots\xmapsto{\alpha_n}\rho_n$ and $\rho_n\Downarrow R'$ with $R'\cap V = R$.
 Since  $\sigma(\gamma) \sqsubseteq \rho$, there exists 
 $\sigma(\gamma)\xmapsto{\alpha_1}\sigma_1\ldots\xmapsto{\alpha_n}\sigma_n$ with $\sigma_i \sqsubseteq \rho_i$ for $i= 1..n$.
 Consequently, $\sigma\tr{\gamma}\sigma(\alpha)\xmapsto{\alpha_1}\sigma_1\ldots\xmapsto{\alpha_n}\sigma_n$ with $\sigma_i \sqsubseteq \rho_i$.
 Hence, there exist $S'\subseteq R'$ such that $\sigma_n\Downarrow S'$. By proposition~\ref{seqofhiddenreductions},
 $\mathcal{A}_V(\sigma) \Downarrow S$ with $S=S'\cap V$. Since $S'\subseteq R'$ we have that $S = S'\cap V \subseteq R'\cap V = R$.
 
 \item Assume $\rho\xmapsto{\alpha}\rho'$. By Proposition~\ref{redabstraccion},   
  $\alpha\in V$ and $\rho' = \bigoplus_{\rho_i\in \mathtt{Alc}(\rho,\alpha,V)} \mathcal{A}_V(\rho_i)$ with
 \[\mathtt{Alc}(\rho,\alpha,V) = \{\rho'\ |\ \rho\xmapsto{\beta_1}\rho_1\ldots\xmapsto{\beta_n}\rho_n\xmapsto{\alpha}\rho'\ \mbox{and}\ \beta_1,\ldots,\beta_n\not\in V\}.\] 
 Since $\sigma(\gamma) \sqsubseteq \rho$, for any $\rho_i$ in $\mathtt{Alc}(\rho,\alpha,V)$ there exists 
 a $\sigma_i$ s.t. $\sigma_i\in\mathtt{Alc}(\sigma(\alpha),\alpha,V)$, i.e., $\sigma(\gamma)\xmapsto{\beta_1}\tau_1\ldots\xmapsto{\beta_n}\tau_n\xmapsto{\alpha}\sigma_i$. 
 Note that $\sigma_i \in\mathtt{Alc}(\sigma(\gamma),\alpha,V)$ implies $\sigma_i \in\mathtt{Alc}(\sigma,\alpha,V)$.
 By proposition~\ref{redabstraccion},
 $\sigma'= \bigoplus_{\sigma_i\in\mathtt{Alc}(\sigma,\alpha,V)} \mathcal{A}_V(\sigma_i)$.
 It remains to show that $(\sigma', \rho')\in \mathcal{S}$. This is done by noting that
 $\sigma' = \bigoplus_j\mathcal{A}_V(\tau_j) \oplus  \tau$ such that any $\tau_j \in \mathtt{Alc}(\sigma,\alpha,V)$ 
 and there is a corresponding $\rho_j$ in $\mathtt{Alc}(\rho,\alpha,V)$ and $\tau_j\sqsubseteq \rho_j$. Note that it can be easily proved that $\sigma_1 \sqsubseteq \rho_1$ and $\sigma_2 \sqsubseteq \rho_2$ implies $\sigma_1\oplus\sigma_2 \sqsubseteq\rho_1\oplus \rho_2$. Consequently,
 $\bigoplus_j \tau_j \sqsubseteq \bigoplus_j\rho_j$. We can easily also prove that $\sigma_1\oplus \sigma_2\sqsubseteq \sigma_1$ for all 
 $\sigma_1, \sigma_2$. Hence, $\bigoplus_j \tau_j \oplus \tau \sqsubseteq \bigoplus_j\rho_j$, and finally, $(\sigma',\rho')\in\mathcal{S}$ by definition of $\mathcal{S}$
\end{enumerate}

\paragraph{Proof of Proposition \ref{prop:abstractionsubcontractvisiblereduction}.}
The proof follows by showing that $\mathcal{S}= \{(\mathcal{A}_V(\sigma)(\gamma),\mathcal{A}_V(\rho)) | \sigma(\gamma) \sqsubseteq \rho$ and $\gamma\in V\}$ 
is a subcontract relation. 
\begin{enumerate}
 \item Assume $\mathcal{A}_V(\rho) \Downarrow R$. By Proposition~\ref{seqofhiddenreductions}, we have that
 $\rho\xmapsto{\alpha_1}\rho_1\ldots\xmapsto{\alpha_n}\rho_n$ and $\rho_n\Downarrow R'$ with $R'\cap V = R$.
 Since  $\sigma(\gamma) \sqsubseteq \rho$, there exists 
 $\sigma(\gamma)\xmapsto{\alpha_1}\sigma_1\ldots\xmapsto{\alpha_n}\sigma_n$ with $\sigma_i \sqsubseteq \rho_i$ for $i= 1..n$.
 Hence, there exist $S'\subseteq R'$ such that $\sigma_n\Downarrow S'$. By proposition~\ref{seqofhiddenreductions},
 $\mathcal{A}_V(\sigma(\gamma)) \Downarrow S$ with $S=S'\cap V$. By Proposition~\ref{prop:abstofvisiblered},  $\mathcal{A}_V(\sigma)(\gamma) = \mathcal{A}_V(\sigma(\gamma))$, hence  $\mathcal{A}_V(\sigma)(\gamma) \Downarrow S$ with $S=S'\cap V$. Since $S'\subseteq R'$ we have that $S = S'\cap V \subseteq R'\cap V = R$.
 
 \item Assume $\rho\xmapsto{\alpha}\rho'$. By Proposition~\ref{redabstraccion},   
  $\alpha\in V$ and $\rho' = \bigoplus_{\rho_i\in \mathtt{Alc}(\rho,\alpha,V)} \mathcal{A}_V(\rho_i)$ with
 \[\mathtt{Alc}(\rho,\alpha,V) = \{\rho'\ |\ \rho\xmapsto{\beta_1}\rho_1\ldots\xmapsto{\beta_n}\rho_n\xmapsto{\alpha}\rho'\ \mbox{and}\ \beta_1,\ldots,\beta_n\not\in V\}.\] 
 Since $\sigma(\gamma) \sqsubseteq \rho$, for any $\rho_i$ in $\mathtt{Alc}(\rho,\alpha,V)$ there exists 
 a $\sigma_i$ s.t. $\sigma_i\in\mathtt{Alc}(\sigma(\gamma),\alpha,V)$, i.e., $\sigma(\gamma)\xmapsto{\beta_1}\tau_1\ldots\xmapsto{\beta_n}\tau_n\xmapsto{\alpha}\sigma_i$. 
  By proposition~\ref{redabstraccion},
 $\sigma'= \bigoplus_{\sigma_i\in\mathtt{Alc}(\sigma(\gamma),\alpha,V)} \mathcal{A}_V(\sigma_i)$. Moreover, $\mathcal{A}_V(\sigma)\tr{\gamma}\sigma'$, i.e. $\mathcal{A}_V(\sigma)(\gamma) = \sigma'$,
 by Proposition~\ref{prop:abstofvisiblered}.
 It remains to show that $(\sigma', \rho')\in \mathcal{S}$. This case follows as for Proposition~\ref{prop:abstractionsubcontracthiddenreduction}.
\end{enumerate}

\paragraph{Proof of Theorem \ref{thm:consistency}.}
We prove by structural induction on  $P$ that all conditions in Definition~\ref{def:consistency} are satisfied.
First of all, note that the language of orchestrators we rely on does not diverge, hence consistency condition (3) is trivially satisfied in all cases
    \begin{itemize}   
      \item ${\bf P = 0}$: Conditions (1), (2) hold trivially since $P$ has no reductions. As far as condition(4) is concerned, note that $\sigma = 0$ and $\sigma \Downarrow R$ implies $R = \emptyset \subseteq A$ for any $A$.  

      \item ${\bf P = \lambda.P'}$. If $\lambda = \tau$ then $P= \tau P'$. The only possible type for $P$ (derived by using rule {\sc (tau)}) is $\sigma$ with $\vdash P': \sigma$ and clearly $\sigma \sqsubseteq \sigma$ and therefore condition (1) holds. Moreover, conditions (2) and (4) trivially hold. Let $\lambda \neq \tau$. Then, condition (1) trivially hold. As regards to condition (2), note that $\vdash P:\sigma$ with $\sigma = \lambda.\sigma'$ and
      $\vdash P' : \sigma'$. Consequently, $\sigma(\lambda)=\sigma'$ and condition (2) holds. As condition (4) is concerned, note that $P\Downarrow R$ implies $R=\{\lambda\} = \{\lambda \ |\ P\tr{\lambda}\}$.

      \item ${\bf P = \mathtt{\Sigma}_{i\in I} \act_i.P_i + \mathtt{\Sigma}_{j \in J} \tau.Q_j}$. From typing rule {\sc (sum)} we have that $\vdash P: \sigma$ with $\sigma = \Sigma_{i \in I} \act_i.\sigma_i \oplus \bigoplus_{j\in J}\rho_j$. Condition (1): If $P \tr{\tau} P'$ then there exists some $k \in J$ such that $P' = Q_k$  and $\vdash P': \rho_k$ with $\vdash Q_k: \rho_k$. Note that $\sigma =  \rho_k \oplus \tau$ for some $\tau$. Consequently, 
      $\sigma = \rho_k \oplus \tau \sqsubseteq \rho_k = \sigma'$. Condition (2): If $P \tr{\lambda} P'$ with $\lambda \neq \tau$, then there exists some $k \in I$ such that $P' = P_k$  and $\act_k = \lambda$ and $\vdash P': \sigma_k$. Consequently, $\sigma(\lambda) = \sigma_k \oplus \tau$ for some $\tau$. Hence, $\sigma(\lambda)\sqsubseteq\sigma'$. As far as condition (4) is concerned, note that $P\centernot{\tr{\tau}}$ implies $J = \emptyset$.Then $\sigma = \Sigma_{i \in I} \act_i.\sigma_i$ then $\sigma\Downarrow R$ implies $R = \{\act_i | i\in I\} =\{ \act \ |\ P \tr{\alpha} \}$.   
         
      \item ${\bf P = P_1|P_2}$. Condition (1), if $P \tr{\tau} P'$  then either $P_1\tr{\tau} P_1'$ or $P_2\tr{\tau} P_2'$. If $P_1\tr{\tau} P_1'$ then $\sigma$ is an internal choice containing a subterm  $\sigma'$ where $\vdash P_1'|P_2 : \sigma'$. Consequently, $\sigma'\sqsubseteq\sigma$. The case $P_2\tr{\tau} P_2'$ is analogous. For condition (2), note that either $P_1\tr{\act} P_1'$ or $P_2\tr{\act} P_2'$ and $\act\neq\tau$. The proof follows as for condition (1). As regards to condition (3), note that neither $P_1\tr{\tau}$ nor $P_1\tr{\tau}$. Hence, $\sigma =  
(\Sigma_{\act_i \in V} \act_i.\sigma_i + \Sigma_{\beta_j \in V} 
   \beta_j.\rho_j)$
 where $P \ltr{\act_i} P_i$ and $Q \ltr{\beta_j} Q_j$
       
    Therefore $\sigma\Downarrow R$ implies $R = \{ \act | P\tr{\act}\}$.
    
    \item ${\bf P = \ifte{m=n} {P_1}{P_2}}$. There are two cases $\opaque \in \{m,n\}$ and $\opaque\not\in \{m,n\}$. Assume 
    $\opaque \in \{m,n\}$. By rule {\sc(cond1)}, $\sigma = \sigma_1 \oplus \sigma_2$ with $\vdash P_1:\sigma_1$ and $\vdash P_2:\sigma_2$. As far as condition (1) is concerned,  $P\tr{\tau} P'$ when either $P_1\tr{\tau} P_1'$ or $P_2\tr{\tau} P_2'$.
    Let  $P_1\tr{\tau} P_1'$ with $\vdash P_1': \sigma_1'$ and $\sigma_1\sqsubseteq \sigma_1'$ by inductive hypothesis. Therefore,
    $\sigma = \sigma_1 \oplus \sigma_2  \sqsubseteq \sigma_1'$. The case $P_2\tr{\tau} P_2'$ follows analogously. 
    For condition (2), the proof follows analogously to condition 1. In respect to condition (3), note that $\sigma\Downarrow R$ implies 
    that either $\sigma_1\Downarrow R$ or $\sigma_2\Downarrow R$. By inductive hypothesis, we know that $\sigma_1\Downarrow R$ implies
    $R\subseteq \{\act |P_1\tr{\act}\}$ and $\sigma_2\Downarrow R$ implies
    $R\subseteq \{\act |P_2\tr{\act}\}$. Hence, $R \subseteq \{\act |P_1\tr{\act}\} \cup \{\act |P_2\tr{\act}\}$. It is easy to see that 
    $P \tr{\act}$ if and only if $P_1\tr{\act}$ or  $P_1\tr{\act}$.
    The cases for 
    $\opaque\not\in \{m,n\}$ follows analogously by noting that $\sigma$ is either $\sigma_1$ or $\sigma_2$ depending on whether $m=n$ or $m\neq n$ hold.
    \end{itemize}

\paragraph{Proof of Proposition \ref{prop:abstractionandreductions}.} 
We only prove the first case above; the second case is similar. 
From $A_V[Q] \tr{\alpha} A_V[Q']$ we have that $Q\tr {\beta} Q'$ with $\beta_{|V} =\alpha$,  being $\beta_{|V}$ defined as the expected counterpart of $\lambda_{|V}$.
By Definition~\ref{def:alt-non-symb-sem},  there exist $M, \lambda, R$ and $\sigma\models M$ 
such that $Q\tr{M,\lambda} R$ and $\lambda\sigma = \beta$ and $R\sigma = Q'$. $P$ and $Q$ are closed, hence
$\bn(\lambda)\cap \fn(P,Q,M) = \emptyset$. Since, $P\propto^V Q$ there exists a $M$-decomposition $D$
such that $\forall M' \in D$, $P\tr{N',\lambda'} P''$ with $M'\Rightarrow N'$, $\lambda_{|V} = \lambda'$,
and there exists some simulation-based abstraction relation  $\mathcal{R}^V_M$ such that 
$P'' \mathcal{R}^{V\cup \vn(\lambda)}_{M'} R$.   Since $\sigma\models M$ and $D$ is a $M$-decomposition, 
there exists at least one $M_i\in D$ such that $\sigma\models M_i$ (and hence $\sigma\models N'$). 
By Definition~\ref{def:alt-non-symb-sem},
$P\tr{\lambda'\sigma} P''\sigma$. There are two cases:
\begin{itemize}
   \item $\lambda' = \tau$ or $\lambda' = \outp{x}{\tuple{a}}$: In both cases, 
   $\lambda$ and $\lambda'$ are closed. Hence, $\lambda' = \lambda'\rho$  
   for any substitution $\rho$. Since, $\beta = \lambda\sigma = \lambda$, we have that $\alpha = \beta_{|V} = \lambda_{|V} = 
   \lambda' = \lambda'\sigma$ .
   
   It remains to show that $P''\sigma\propto^V Q'$ with $Q' = R\sigma$. Since $\bn(\lambda) = \emptyset$, we have $P'' \mathcal{R}^{V}_{M'} R$.
   Also note that $R$ and $P''$ are closed because $Q$ and $P$ are closed. Hence, $Q' = R\sigma = R$, $P'' = P''\sigma$     and $P'' \sigma \mathcal{R}^{V}_{M'} Q'$.
   %Since $P''\sigma$ and $Q'$ are closed, $P''\sigma \propto^{V} R\sigma$ holds by Proposition~\ref{prop:condition-cut}.
   \item $\lambda' = \inp{x}{\tuple{v}}$:  Since $\lambda' = \lambda_{|V}$ is an input action, we have that $\lambda$ is an input action and 
   both $\lambda'$ and $\lambda$ have the same subject $x$ that belongs to $V$. Consequently,
   $\lambda_{|V} = \lambda$. Then, $\lambda' = \lambda_{|V} = \lambda$, and consequently
   $\lambda'\sigma = \lambda \sigma = \beta$. Since, $\beta$ is an input action whose subject is in $V$, $\beta_{|V} = \beta$. Consequently,
   $\lambda'\sigma = \lambda\sigma = \beta_{|V} = \alpha$.
   It remains to show that $P''\sigma \propto^V Q'$ with $Q' = R\sigma$. We know that $P'' \mathcal{R}^{V\cup \vn(\lambda)}_{M'} R$. Since
   $\lambda$ is an input action $\tuple{v}\cap\fn(M') =\emptyset$, hence $\sigma\models M'$. By Proposition~\ref{prop:visible-cut},
   $P''\sigma \mathcal{R}^{V}_{M'} R\sigma$. Since $P''\sigma$ and $R\sigma$ are closed, $P''\sigma \propto^{V} R\sigma$ holds by Proposition~\ref{prop:condition-cut}.
\end{itemize}

\end{document}